\documentclass[notitlepage,a4,10.5pt]{article}

\usepackage[backend=biber,style=alphabetic,doi=true,url=false]{biblatex}
\usepackage[colorlinks=true,linkcolor=blue,citecolor=blue,urlcolor=blue]{hyperref}

\usepackage{array}
\usepackage{booktabs}  
\usepackage{multirow}  
\usepackage[margin=1in]{geometry}

\usepackage{graphicx}

\usepackage{amsmath}%
\usepackage{amsfonts}%
\usepackage{amssymb}%
\usepackage{mathrsfs}
\usepackage{amsthm}
\usepackage{float}

\usepackage{authblk}
\usepackage{enumitem}

\usepackage{xcolor}
\usepackage{appendix}

\newcommand{\mc}{\mathcal}

\newcommand{\bol}{\boldsymbol}

\newcommand{\abs}[1]{\left\lvert{#1}\right\rvert}
\newcommand{\w}{\wedge}
\newcommand{\lr}[1]{\left({#1}\right)}
\newcommand{\lrs}[1]{\left[{#1}\right]}
\newcommand{\lrc}[1]{\left\{{#1}\right\}}

\newcommand{\mf}{\mathfrak}
\newcommand{\p}{\partial}

\newcommand{\ti}[1]{\textit{#1}}

\newtheorem{mydef}{\textit{Definition}}[section]
\newtheorem{remark}{\textit{Remark}}[section]

\newtheorem{proposition}{\textit{Proposition}}[section]

\newcommand{\eq}[1]{\begin{equation}\begin{split}{#1}\end{split}\end{equation}}
\newcommand{\sys}[2]{\begin{subequations}\begin{align}{#1}\end{align}\label{#2}\end{subequations}}

\begin{document}

\title{
A Collision Operator for Field-Mediated Interactions in General Relativistic Kinetic Theory
}
\author[1,2]{Naoki Sato} 
\affil[1]{National Institute for Fusion Science, 
322-6 Oroshi-cho Toki-city, Gifu 509-5292, Japan }
\affil[2]{Graduate School of Frontier Sciences, The University of Tokyo, 
Kashiwa, Chiba 277-8561, Japan
\protect\\ Email: sato.naoki@nifs.ac.jp}
\date{\today}
\setcounter{Maxaffil}{0}
\renewcommand\Affilfont{\itshape\small}

    \maketitle
    \begin{abstract}  
   We develop a Hamiltonian framework for general relativistic kinetic theory on the cotangent bundle $T^*M$ of a Lorentzian (pseudo-Riemannian) manifold. Starting from the geodesic Hamiltonian $H$, we derive a Landau-type collision operator for self-gravitating particles undergoing binary interactions mediated by an arbitrary potential energy $V$, and couple the resulting kinetic stress--energy to the Einstein field equations to obtain the Landau–Einstein system. In the presence of a coordinate-time Killing symmetry we find a family of stationary states of the form $f \propto \gamma \exp[-\beta(H+\Phi)]\,\zeta(p_0)$, where $\Phi$ is the mean field, $\gamma=dt/d\tau$, $\beta$ is an inverse-temperature parameter, and $\zeta$ encodes symmetry-induced degeneracy. 
\end{abstract}


\tableofcontents

\section{Introduction}


The collective behavior of massive bodies---such as high-energy astrophysical systems, gravitationally bound systems in early-universe cosmology, or self-gravitating particle ensembles---is governed by relativistic kinetic theory. 
This framework couples a kinetic equation for the system's distribution function with a self-consistently generated spacetime, whose curvature is determined by Einstein's field equations. 
In its standard form, this coupling provides a unified description of matter and geometry across a wide range of relativistic systems.

Nevertheless, an important aspect of relativistic kinetic theory has not yet been fully explored. 
Standard formulations rely on the assumption that binary collisions occur in a locally Minkowskian spacetime, 
where the four-momentum is treated as an additive invariant. 
Although this simplification is well justified in many practical settings, 
where the localized nature of collisions renders curvature effects negligible during individual scattering events, 
there may exist situations in which 
this assumption is not generally compatible with strong curvature---as, for instance, in the vicinity of a black hole---or with interactions governed by long-range forces. 
Our purpose here is to demonstrate that this assumption carries deeper implications for the very notions of thermodynamic equilibrium, 
temperature, and entropy in general relativity, even when the approximation itself remains physically consistent.

The reason is that the additive invariants of collisions determine the structure of statistical steady states, 
and thereby the corresponding thermodynamics. 
When the usual assumption is relaxed, a genuinely general relativistic thermal distribution should emerge---one that cannot be characterized in terms of 
special relativistic or Newtonian invariants.




The broader context of this study will be elaborated in Section~2. 
Briefly, we formulate a general‐relativistic kinetic theory on the cotangent bundle $T^*M$ of a Lorentzian manifold $(M,g)$. 
Exploiting the Hamiltonian structure of the geodesic Hamiltonian, we construct a Landau–type collision operator. 
Our main results are: (i) the \ti{Landau--Einstein system}, and (ii) its thermodynamic equilibrium, the \ti{Landau--Einstein distribution}.
These names are adopted to emphasize both the kinetic character of the equation (of Landau type) and its coupling to the Einstein field equations, in analogy with the Vlasov--Maxwell or Vlasov--Einstein systems.

Let $\lr{M, g}$ denote a $4$-dimensional Lorentzian manifold with coordinates $x = \lr{x^0, x^1, x^2, x^3}$. We denote by $T_xM$ the tangent space at $x$, by $T_x^{\ast}M$ the cotangent space at $x$, 
by $p = \lr{p_0, p_1, p_2, p_3} \in T^{\ast}_x M$ the $4$-momentum, by $z=\lr{x,p}$ the phase space coordinates, and by $dz=dx^0dx^1dx^2dx^3dp_0dp_1dp_2dp_3$ the phase space volume element. 
We adopt the signature convention $\lr{-, +, +, +}$. Greek letters will be used for spacetime indices, lower case  Roman letters will be reserved for spatial indices, and indices involving  both spacetime and momentum will be specified with upper case Roman letters. We also employ the Einstein summation convention over repeated indices.
The Lorentz factor is defined as $\gamma = \frac{dt}{d\tau}$,  
where $t=x^0/c$ is the coordinate-time, $c$ is the speed of light, and $\tau$ is the proper time. 
We denote by $G$ the gravitational constant, by $R^{\mu\nu}$ the Ricci curvature tensor, and by $R = g_{\mu\nu} R^{\mu\nu}$ the Ricci scalar. 
We further assume the minimal hypotheses ensuring that  integrals are well defined—namely, that the relevant manifolds carry a smooth positive density (e.g., the Liouville measure on $T^*M$) and that the fields have sufficient regularity for the constructions of the collision operator.  
Let $f(z)$ denote the distribution function of an ensemble of $N$ particles of mass $m$, with normalization 
$N = \int f\, dx^1 dx^2 dx^3 dp_0 dp_1 dp_2 dp_3$. 

\begin{mydef}
The Landau--Einstein system for the unknowns $\lr{f,g}$ is 
\begin{subequations} \label{LE}
\begin{align}
&\lrc{\frac{f}{\gamma},\mc{H}} = \mathscr{C}\lr{f,f}, \\
&R^{\mu\nu} - \frac{1}{2} R g^{\mu\nu} = \kappa T^{\mu\nu}, 
\end{align}
\end{subequations}
where
\eq{
\mc{H} = \frac{1}{2}mc^2\lr{1+\frac{g^{\mu\nu}p_{\mu}p_{\nu}}{m^2c^2}} {+\Phi},\label{gH}
}
is the single-particle Hamiltonian function,
\eq{
\Phi = {\int f\lr{z'} V\lr{z,z'}\, dz'},\label{Phi}
}
is the {ensemble-averaged interaction potential energy}  associated with the binary interaction potential $V \lr{z,z'}$,
\eq{
\lrc{F, G} = \frac{\p}{\p z^A} \lr{ {J_c}^{AB} \frac{\p G}{\p z^B} F },\label{cPB}
}
is the canonical Poisson bracket on smooth functions $F\lr{z}, G\lr{z}$ with canonical Poisson tensor $J_c$,
\eq{
\mathscr{C}\lr{f,f} = \frac{\p}{\p z^A} \lrc{ \frac{f}{\gamma} J_{\Sigma}^{AB} \int \frac{f'}{\gamma'} \lrs{ \tilde{{\Pi}}_{\mc{V}BC} J_{\Sigma}'{}^{CD} \frac{\p \log \lr{f'/\gamma'}}{\p z^D} - \Pi_{\mc{V}BC} J_{\Sigma}^{CD} \frac{\p \log \lr{f/\gamma}}{\p z^D} }\, dz' },\label{LEC}
}
is the Landau--Einstein collision operator, where $'$ denotes evaluation at $z'$.
The tensor 
$J_{\Sigma}$ is the projection of the canonical Poisson tensor $J_c$ on the submanifold of constant coordinate-time and constant light speed, the latter given by the constraint $\chi=c^2+g_{\mu\nu}\dot{x}^{\mu}\dot{x}^{\nu}=0$, where the dot indicates differentiation with respect to proper time.  
The interaction tensors $\tilde{{\Pi}}_{\mc{V}}\lr{z,z'}$ and $\Pi_{\mc{V}}\lr{z,z'}$ are determined by $V$.
\eq{
\kappa = \frac{8\pi G}{c^4},
}
is a positive real constant, and $T^{\mu\nu}$ is the energy--momentum tensor
\eq{ 
T^{\mu\nu} = \int \frac{f\, p^{\mu} p^{\nu}}{m\gamma \sqrt{-\mf{g}}} \, dp_0 dp_1 dp_2 dp_3 +{F^{\mu\nu}+} C^{\mu\nu},\label{EMT}
}
where $\mf{g}$ is the determinant of the metric tensor, 
$F^{\mu\nu}$ is the energy-momentum associated with the potential energy $\Phi$, 
and the symmetric tensor $C^{\mu\nu}$ accounts for collisions.
\end{mydef}
Further details on $V$, $J_{\Sigma}$,  $\tilde{\Pi}_{\mc{V}AB}$, $\Pi_{\mc{V}AB}$, $F^{\mu\nu}$, and $C^{\mu\nu}$ will be provided later. Here, we remark that the integrals appearing in \eqref{Phi} and \eqref{LEC} include integrations with respect to coordinate-time, and that both $V$ and the integrands vanish whenever $z$ 
and $z'$ are not causally related. This ensures that the interaction respects causality by accounting for the finite speed of propagation.
 
We will show that, depending on the physical assumptions imposed on the collision process, the Landau--Einstein system~\eqref{LE} admits thermodynamic equilibrium states; in particular, we identify two families: the Landau--Synge and the Landau--Einstein equilibria. 
\begin{mydef}
The {Landau–Synge distribution} is the coordinate-time-independent distribution function
\begin{equation}
f_{\infty}\lr{x^1,x^2,x^3,p_0,p_1,p_2,p_3} = \frac{1}{Z} \, \gamma\,\delta\lr{\chi 
}\,
\exp\lrc{
\beta^{\mu} p_{\mu}}
,
\label{LSd}
\end{equation}
where  
$\mc{H}$, $g$, and $\Phi$  possess a coordinate-time Killing symmetry,
$\beta^\mu\in\mathbb{R}$ is a timelike future-directed vector, and $Z$ a normalization constant. 
\end{mydef}
We note that, for $\Phi=0$ and flat spacetime $g_{\mu\nu}=\eta_{\mu\nu}$ (with $\eta$ the Minkowski metric), the kinematics gives
$p^{\mu}=m\,{dx^{\mu}}/{d\tau}$, and hence \eqref{LSd} reduces to the Synge--J\"uttner distribution.
The factor $\gamma$ arises  from using the coordinate-time $t$ for bookkeeping rather than the proper time $\tau$ (i.e., $dt=\gamma\,d\tau$). 

\begin{mydef}
The {Landau--Einstein} distribution is the coordinate-time-independent distribution function 
\eq{
f_{\infty}\lr{x^1,x^2,x^3,p_0,p_1,p_2,p_3}=\frac{1}{Z}\gamma\,\delta\lr{\chi} \,\exp\lrc{-\beta \mc{H}}\,\zeta\lr{p_0},\label{LEd}
}
where $\mc{H}$, $g$, and ${\Phi}$ possess a coordinate-time Killing symmetry,   $\beta\in\mathbb{R}$, $Z$ is a normalization constant, and $\zeta$ a function of $p_0$. 
\end{mydef}
\begin{remark}
The function $\zeta(p_{0})$ parametrizes the degeneracy associated with 
additional collisional invariants involving the coordinate-time momentum $p_{0}$.  
If binary interactions conserve only the pair   energy $\mc{H}+\mc{H}'$, then no such 
degeneracy arises and one may set $\zeta= 1$ without loss of generality.  
We retain $\zeta$ in the general definition in order to accommodate situations 
in which $p_{0}$ (or $p_{0}+p_{0}'$) is also an invariant of the collision 
process, in which case $\zeta$ is fixed (up to normalization) by the additional 
symmetry; for instance, if $p_{0}+p_{0}'$ is additive, then 
$\zeta(p_{0})\propto e^{\beta^0 p_{0}}$.  
See Section~6.2 for further discussion.
\end{remark} 
We will show that both \eqref{LEd} and \eqref{LSd} represent distinct manifestations of a common underlying structure of general relativistic equilibria (see Sec.~6 and Fig.~2 below), 
which encodes information about the bookkeeping time parameter, the speed-of-light constraint, and the associated collisional invariants.


The present manuscript is organized as follows. 
Section~2 is a preliminary section where we discuss the main issues that separate classical and general relativistic kinetic theory (including the notions of thermodynamic equilibrium and temperature, 
conservation of energy and momentum, the existence of an invariant measure, and the formulation of the first and second laws of thermodynamics), and review the relevant literature.
Section~3 provides a Hamiltonian formulation of binary collisions in a general relativistic spacetime driven by a general interaction potential energy. 
In Section~4, we adopt this description of binary scatterings to derive a general relativistic Landau-type collision operator. 
The resulting Landau--Einstein system, its conservation laws,  and the H-theorem are discussed in Section~5, while the properties of the thermodynamic equilibrium states and their relationship with the special relativistic and classical counterparts are described in Section~6. 
The notion of relativistic temperature emerging from this construction is examined in Section~7, while concluding remarks are given in Section~8.


\section{Structural differences between classical and general relativistic kinetic theory}

In this section, we focus on the fundamental issues that arise in the formulation of a general relativistic kinetic theory. 
We identify the structural differences between classical,
special relativistic, and general relativistic kinetic theory that motivate 
the construction developed in later sections. 
Our discussion has three goals:
(i) to clarify the various notions of equilibrium available in a relativistic 
setting and their classical and Minkowski limits;
(ii) to explain why the usual additive invariants $p_{\mu}+p'_{\mu}$ used in 
special relativistic kinetic theory fail to generalize covariantly in curved 
spacetime, motivating the use of the  Hamiltonian $\mc{H}$;
(iii) to address the role of invariant measures and entropy in a 
coordinate-time description.  
These considerations justify the structural assumptions underlying the 
Landau--Einstein collision operator developed in Sections 3--5.

We refer the reader to \cite{MTW,Cardenas,Sarbach,Sarbach14} for comprehensive treatments of general relativistic kinetic theory within the framework of differential geometry. 
The subject remains an active area of research, motivated by its importance in astrophysical applications (e.g., \cite{Gab1,Gab2}). 
The standard approach to relativistic kinetic theory is typically formulated in terms of a collisionless (Vlasov-type) model or, when collisions are included, assumes \ti{special relativistic elastic} scattering processes, leading to the Synge-J\"uttner  distribution~\cite{Israel,Cardenas,Sarbach,Sarbach14}, \cite[pp.~6--8]{Hakim}. 
We will review the details of these constructions later; for the moment, it suffices to note that such formulations, while internally consistent and physically well-founded, are naturally confined to collisions taking place in flat spacetime. 
This restriction, in turn, specifies the class of admissible thermodynamic equilibria and identifies $p_0$---the Noether invariant associated with the coordinate-time Killing symmetry of the geodesic Hamiltonian---as the conserved energy during binary interactions, with the spatial components $p_i$ ($i=1,2,3$) serving as additional additive invariants. 

In what follows, we explore how this framework may be extended to a fully general relativistic setting, in which the geodesic Hamiltonian itself is regarded as the `physical' energy and particles are allowed to deviate from geodesic motion during interactions. 
A schematic representation of this extension---contrasting the standard collision model with the scattering framework developed here---is shown in Fig.~1. 
Before proceeding, however, it is necessary to clarify what is meant by thermodynamic equilibrium in a general relativistic context.

\begin{figure}
  \centerline{\includegraphics[scale=0.25]{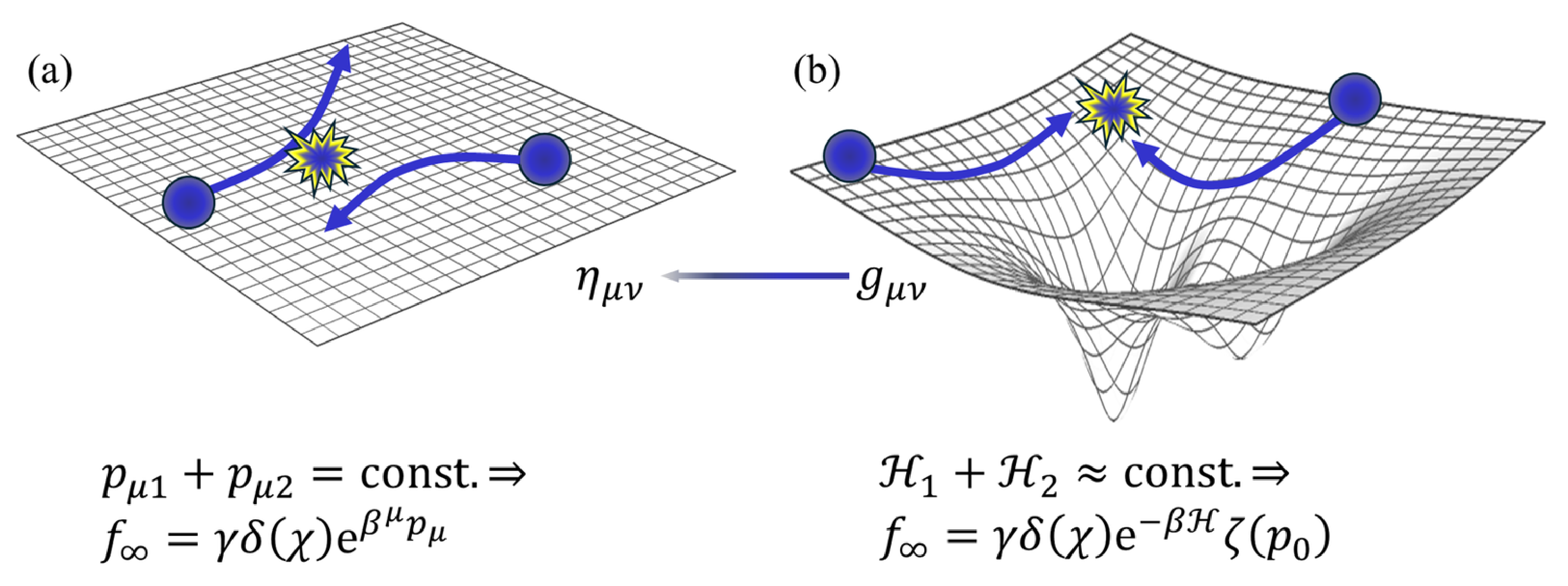}}
  \caption{ Schematic comparison between (a) the standard special-relativistic scattering process  where the components $p_{\mu}$ of the four-momentum are additive invariants yielding the Synge–J\"uttner distribution, and (b)  the general-relativistic collision model developed here, where the geodesic Hamiltonian $\mc{H}$, including mean-field contributions, is an approximate additive  invariant yielding the Landau–Einstein distribution.}
\label{fig1}
\end{figure}

\subsection{Thermodynamic equilibrium and temperature}
Due to the relativistic nature of time, the notion of thermodynamic equilibrium---which is typically associated with a non-evolving or steady state---becomes nontrivial in a relativistic setting. 
This issue arises already in the flat Minkowski spacetime of special relativity, where the absence of a preferred global time coordinate leads to various alternative notions of equilibrium. 

To illustrate this, let us consider a distribution function $f(x, p, \tau)$ that depends explicitly on the proper time $\tau$, 
which serves as the Hamiltonian evolution parameter governing the dynamics of the phase-space coordinates $\lr{x, p}$. 
In this study, we restrict our attention to distribution functions that are stationary in proper time:
\begin{equation}
\frac{\partial f}{\partial \tau} = 0.
\end{equation}
Such distributions, which we henceforth denote simply as $f(x,p)$, may be viewed as a form of \textit{spacetime equilibrium}, being steady with respect to proper time.  
However, this condition is somewhat limited, since the coordinate-time $t = x^0 / c$ still appears explicitly in the arguments of $f(x,p)$,  allowing for evolution when viewed from a coordinate-time observer.

In this work, we do not explore the physical implications of proper-time-dependent distribution functions---although such generalizations may be relevant in other contexts. We simply observe that using proper time as a common Hamiltonian evolution parameter for an ensemble of particles is mathematically consistent and does not conflict with its physical interpretation as a quantity distinct from coordinate-time~$t$.

Let $\mc{H}\lr{x,p}$ denote the Hamiltonian of a single particle, and suppose that particle interactions are limited to binary collisions. Then, conservation of phase space volume, and the independence of $f$ from $\tau$, leads to the following kinetic equation for $f$, 
\eq{
\lrc{f,\mc{H}}
=\lr{\frac{df}{d\tau}}_{\rm coll},\label{dfdt0}
}
where, as usual, $\lrc{\cdot,\cdot}$ is the canonical Poisson bracket \eqref{cPB}, and the term on the right-hand side accounts for collisions. 
In classical kinetic theory~\cite{KampenBBGKY,Marsden},  \cite[pp. 94-127]{Hakim}, 
such collision term arises 
as an approximation to the coupling between the 
first and second equations of the 
BBGKY hierarchy, which captures the statistical correlations encoded in the two-particle distribution function 
$f_2\lr{x_1,p_1,x_2,p_2}$, where $\lr{x_i,p_i}$ are the phase space coordinates of particle $i=1,2$. 
In general, $f_2$ is not the product of the single-particle distributions, i.e., $
f_2\lr{x_1,p_1,x_2,p_2} \neq f\lr{x_1,p_1}f\lr{x_2,p_2}$.
Physically, the collision term $\lr{df/d\tau}_{\rm coll}$ is responsible for entropy production (reflecting the information loss due to closure of the BBGKY hierarchy), and thus for the thermalization of the system. One natural definition of thermodynamic equilibrium is that such thermalized, maximum entropy states (see sec 2.3 for further details on the relativistic notion of entropy) are characterized by the vanishing of the collisional flux:
\eq{
\lr{\frac{df}{d\tau}}_{\rm coll} = 0 \iff \lrc{f,\mc{H}} = 0.\label{Syeq}
}
We refer to this condition as a collisional  \ti{Liouville equilibrium}, since the vanishing of collisional effects reduces the system to Liouville’s setting, where the distribution is constant along Hamiltonian trajectories. 
We emphasize that such configurations are distinct from solutions of the Vlasov equation $\lrc{f,\mc{H}}=0$ \cite{Kandrup}, as the collision term must vanish \textit{in addition}  to satisfying the Hamiltonian flow condition.

This perspective underpins Synge’s relativistic kinetic theory~\cite{Israel,Weert,Ehlers}, where equilibrium is described via the vanishing divergence of the entropy flux vector: $\nabla_{\mu}S^{\mu} = 0$, with $\nabla_{\mu}$ denoting the covariant derivative. We refer to this theoretical framework as \textit{Synge theory}, and the physical system it describes as a \textit{Synge gas}. Synge theory has been applied to model irreversible processes in relativistic hydrodynamics~\cite{Israel2,Israel3,Israel4}. 

Synge theory describes transport processes in a relativistic gas, neglecting quantum effects and all non-gravitational interactions except for binary elastic collisions. Between collisions, particle worldlines follow geodesics. Collisions are modeled as instantaneous events occurring in locally Minkowskian frames, where gravitational effects are negligible. In such frames, the total momentum of the colliding particles is conserved. In this context, the kinetic equation \eqref{dfdt0} reduces to the relativistic Boltzmann equation~\cite{Israel,Walker}:
\begin{equation}
\frac{p^{\mu}}{m}\frac{\partial f}{\partial x^{\mu}} = \left(\frac{df}{d\tau}\right)_{\mathrm{coll}},
\end{equation}
with the collision operator given by
\begin{equation}
\left(\frac{df}{d\tau}\right)_{\mathrm{coll}} = \frac{1}{m} \int W(p^{\ast}, p^{\ast\prime}; p, p^\prime) 
\left[ f(x, p^{\ast}) f(x, p^{\ast\prime}) - f(x, p) f(x, p^\prime) \right] \, d\omega' d\omega^{\ast} d\omega^{\ast\prime},\label{CSy}
\end{equation}
where \( W(p^{\ast}, p^{\ast\prime}; p, p^\prime) \) is a symmetric scattering kernel describing the transition from initial states \( p, p^\prime \) to final states \( p^{\ast}, p^{\ast\prime} \). The integral is over solid angles $\omega$. In a locally Minkowskian frame \( d\omega = dp_1 dp_2 dp_3 / (-n_{\mu} p^{\mu}) \), with $n^{\mu}$ the normal of a three-dimensional surface $dS=dxdydz$.  

In general, the equilibrium condition \eqref{Syeq} implies that \( f \) remains constant along particle worldlines, with the thermodynamic arrow of time set by proper time \( \tau \) (see section 2.3 for further details on the notion of thermodynamic arrow of time). This differs fundamentally from equilibrium defined by the existence of a coordinate-time  symmetry, i.e., \( \partial f / \partial t = 0 \), which characterizes steady states relative to a particular observer. 
Indeed, \eqref{Syeq} does not imply $\p f/\p t=0$. 

Now consider an observer in a coordinate system $x = \lr{x^0,x^1,x^2,x^3}$.
Such an observer may define equilibrium as a state for which physical  observables, including the spacetime metric $g$, are time-independent in coordinate-time $t = x^0/c$, i.e., $\p/\p t=0$. 
This corresponds to the existence of a Killing symmetry with respect to coordinate-time. We call any state satisfying 
\eq{
\frac{\p f}{\p t}=0,~~~~\frac{\p g}{\p t}=0,\label{Keq}
}
a \ti{Killing equilibrium}. This paper focuses on Killing equilibria, which we will refer to simply as \ti{thermodynamic equilibria} if, in addition to \eqref{Keq}, the collision term vanishes, $\lr{df/d\tau}_{\rm coll}=0$.

In what follows, we call the limit $g \rightarrow \eta$ the \ti{Minkowski limit} and 
the limit $c \rightarrow +\infty$ the \ti{classical limit}. 
Spacetime equilibria become trivial in the classical limit, while  
Killing equilibria recover the classical notion of thermodynamic equilibrium in the same limit. The Minkowski and classical limits of Liouville equilibria can be obtained in terms of the condition that the collision term in the relevant limit of the kinetic equation 
\eqref{dfdt0} vanishes. To see this, it is convenient to define
the $p_0$-averaged distribution function 
\eq{
\mf{f}\lr{t, \bol{x}, \bol{p}} = \int f\, dp_0,\label{fp0}
}
where $\bol{x} = \lr{x^1, x^2, x^3}$ and $\bol{p} = \lr{p_1, p_2, p_3}$.  
In the Minkowski limit, integrating with respect to $p_0$ and dropping boundary terms, 
equation \eqref{dfdt0} becomes
\eq{
\frac{\p}{\p t} \int \gamma f\, dp_0 + \lrc{\mf{f}, \mc{H}_{\rm cl}}_{\rm s} = \lr{\frac{d\mf{f}}{d\tau}}_{\rm coll}, 
}
where 
\eq{
\lr{\frac{d\mf{f}}{d\tau}}_{\rm coll} = \int \lr{\frac{df}{d\tau}}_{\rm coll}\, dp_0,
} 
$\mc{H}_{\rm cl}$ is the classical particle energy 
\eq{
\mc{H}_{\rm cl} = \frac{\bol{p}^2}{2m} + \Phi, 
}
with $\bol{p}^2=p_i\eta^{ij}p_j$, 
and $\lrc{\cdot, \cdot}_{\rm s}$ is the spatial canonical Poisson bracket 
\eq{
\lrc{F, G}_{\rm s} = \frac{\p F}{\p x^i} \frac{\p G}{\p p_i} -
\frac{\p G}{\p x^i} \frac{\p F}{\p p_i}.
}
If one further assumes that
that $\Phi$ does not depend on $\bol{p}$ and that particles are always `on-shell', 
$\gamma = \sqrt{1 + \bol{p}^2 / m^2 c^2}$. 
This leads to the special-relativistic kinetic equation 
\eq{
\gamma\frac{\p\mf{f}}{\p t} + \lrc{\mf{f}, \mc{H}_{\rm cl}}_{\rm s} = \lr{\frac{d\mf{f}}{d\tau}}_{\rm coll}, \label{fM}
}
with Liouville equilibrium 
\eq{
\lr{\frac{d\mf{f}}{d\tau}}_{\rm coll} = 0 \iff \frac{\p}{\p t} \lr{\gamma \mf{f}} + \lrc{\mf{f}, \mc{H}_{\rm cl}}_{\rm s} = 0.  
}
If we further take the classical limit of \eqref{fM}, we obtain 
\eq{
\frac{\p \mf{f}}{\p t} + \lrc{\mf{f}, \mc{H}_{\rm cl}}_{\rm s} = \lr{\frac{d\mf{f}}{dt}}_{\rm coll},
}
with
\eq{
\lr{\frac{d\mf{f}}{dt}}_{\rm coll} = \lim_{c \rightarrow +\infty} \lr{\frac{d\mf{f}}{d\tau}}_{\rm coll}.
}
The corresponding Liouville equilibrium is 
\eq{
\lr{\frac{d\mf{f}}{dt}}_{\rm coll} = 0 \iff \frac{\p \mf{f}}{\p t} + \lrc{\mf{f}, \mc{H}_{\rm cl}}_{\rm s} = 0. 
}
These equilibrium notions and their Minkowski and classical limits are summarized in Table~1. 
The hierarchy in Table~1 highlights that these equilibrium notions are
logically distinct.  
Spacetime equilibrium ($\partial_{\tau} f=0$) is a purely kinematic 
condition reflecting the use of proper time as Hamiltonian parameter.
Liouville equilibrium requires, in addition, the vanishing of collisional 
effects and therefore characterizes the natural states of the 
collisionless flow.
Killing equilibrium, by contrast, is an observer-based notion tied to the 
existence of a coordinate-time symmetry and is the relevant condition for 
thermodynamic equilibrium in general relativity.
In this work, ``thermodynamic equilibrium'' always refers to the 
simultaneous satisfaction of Killing symmetry and vanishing collision term. 
\begin{table}[h!]
\centering
\renewcommand{\arraystretch}{2.1} 
\setlength{\tabcolsep}{10pt}

\begin{tabular}{>{\bfseries}l c c c}
\toprule
Equilibrium & \textbf{Condition} & \textbf{Minkowski limit} & \textbf{Classical limit}  \\
\midrule

\textbf{Spacetime} &
$\frac{\p f}{\p\tau}=0$ &
$\frac{\p f}{\p\tau}=0$ &
$\mf{f}\!\lr{t,\bol{x},\bol{p}}$ \\
[12pt]

\textbf{Liouville} &
$\lr{\frac{df}{d\tau}}_{\rm coll} = \lrc{f,\mc{H}} = 0$ &
$\begin{aligned}[t]
&\lr{\frac{d\mf{f}}{d\tau}}_{\rm coll} = \\[-2pt]
&\gamma\frac{\p\mf{f}}{\p t}+\lrc{\mf{f},\mc{H}_{\rm cl}}_{\rm s} = 0
\end{aligned}$ &
$\begin{aligned}[t]
&\lr{\frac{d\mf{f}}{dt}}_{\rm coll} = \\[-2pt]
&\frac{\p\mf{f}}{\p t} + \lrc{\mf{f},\mc{H}_{\rm cl}}_{\rm s} = 0
\end{aligned}$ \\
[14pt]

\textbf{Killing} &
$\frac{\p f}{\p t}=0,\ \frac{\p g}{\p t}=0$ &
$\frac{\p f}{\p t}=0,\ g=\eta$ &
$\frac{\p\mf{f}}{\p t}=0$ \\
[12pt]

\textbf{Thermodynamic} &
$\frac{\p f}{\p t}=\lr{\frac{df}{d\tau}}_{\rm coll}=0,\ \frac{\p g}{\p t}=0$ &
$\frac{\p f}{\p t}=\lr{\frac{df}{d\tau}}_{\rm coll}=0,\ g=\eta$ &
$\frac{\p\mf{f}}{\p t}=\lr{\frac{d\mf{f}}{dt}}_{\rm coll}=0$ \\
\bottomrule
\end{tabular}

\caption{Notions of general relativistic collisional equilibrium for the distribution functions $f$ and the metric tensor $g$, along with their Minkowski and classical limits. Note that Liouville, Killing, and thermodynamic equilibria are all spacetime equilibria since it is assumed that $\p f/\p\tau =0$.}
\label{tab:tab1}
\end{table}




As it will be discussed in Section~7, the notion of temperature
acquires a different meaning depending on the adopted notion of equilibrium.

\subsection{Energy and momentum}
The essential nature of a kinetic theory lies in the way 
collisions are modeled. 
Indeed, collisions are the mechanism by which 
a statistical ensemble may access a thermalized state that is (at least partially) independent of its initial configuration, 
characterized by a finite set of parameters (e.g., temperature), and relevant from an experimental standpoint. 
In the classical Boltzmann transport equation for a dilute gas~\cite[pp. 60-62]{Huang}, collisions are modeled as binary interactions where gas molecules physically collide. 
Synge's collision operator~\eqref{CSy} generalizes the Boltzmann collision operator to a special-relativistic setting.  
In the Landau collision operator~\cite{Landau1936,KampenBBGKY,Thompson,Fitz} for grazing Coulomb collisions in a plasma, binary interactions are mediated by the long-range Coulomb potential. 

These kinetic models share the assumption that collisions are \ti{elastic}, and the derivation of the scattering kernel (or differential cross-section) relies on the conservation of energy and momentum. 
Specifically, each component of the $4$-momentum is treated as an \ti{additive invariant} during binary collisions:
\eq{
p_{\mu 1} + p_{\mu 2} = p_{\mu 1}' + p_{\mu 2}',
}
where $p_{\mu i}$ denotes the $4$-momentum of particle $i=1,2$ before the collision, and $p_{\mu i}'$ after. 
The existence of these invariants has deep consequences for the accessible steady states \cite{Cercignani}: since collisions do not alter the total $4$-momentum, it is not surprising that the statistical  equilibria  
include the exponential weight 
\eq{
e^{\beta^{\mu}p_{\mu}}.\label{fpmu}
}

The conservation laws above ultimately trace back to symmetries of the Hamiltonians governing the dynamics and interactions. 
For example, in the relativistic Boltzmann case, the energy of a particle during a short-range elastic interaction may be expressed as
\eq{
h_i = -cp_{0i}+V\lr{x_1 - x_2, \bol{p}_1, \bol{p}_2} = mc^2\sqrt{1 + \frac{\bol{p}_i^2}{m^2c^2}} + V\lr{x_1 - x_2, \bol{p}_1, \bol{p}_2}, \qquad i=1,2,
}
where $V$ is a short-range, causal scattering potential. 
In the Landau model, the energy is instead given by
\eq{
h_i = \frac{\bol{p}^2_i}{2m} + V\lr{|\bol{x}_1 - \bol{x}_2|}, \qquad i=1,2,
}
where $V$ is the Coulomb potential. 
In both cases, the rate of change in total momentum during interaction vanishes:
\eq{
\dot{p}_{\mu 1} + \dot{p}_{\mu 2} = -\frac{\p h_1}{\p x^{\mu}_{1}} - \frac{\p h_2}{\p x^{\mu}_{2}} = 0.
}

However, when we move to a general relativistic setting, these additive invariants no longer exist. 
In curved spacetime, the functions $h_i$ do not represent Hamiltonians, and the notion of elastic scattering loses its standard meaning. 
What is typically considered as “physical energy” in special relativity— the $0$th component $p_0$—is no longer conserved unless the metric admits a coordinate-time Killing symmetry. 
Indeed, for the Minkowski metric $g = \eta$, the geodesic Hamiltonian
\eq{
{H}_{\eta} = \frac{1}{2} mc^2\left(1 + \frac{\eta^{\mu\nu}p_{\mu}p_{\nu}}{m^2c^2}\right)
}
implies 
\eq{
\dot{p}_0 = -\frac{\p {H}_{\eta}}{\p x^0} = 0,
}
but in a general curved geometry $g\neq \eta$, this symmetry and conservation law no longer hold. We also emphasize that $p_0$ is not a scalar, while energy is. 

One might argue that collisions occur over sufficiently small scales to be modeled locally in flat spacetime. 
While this is indeed appropriate in many situations, there may exist regimes in which the interaction range becomes comparable to the local curvature scale---for instance, in the vicinity of a black hole---so that this local approximation no longer applies straightforwardly. 
More generally, it is worth noting that the equilibrium form \eqref{fpmu} arises from a symmetry leading to additively  conserved momenta, rather than from a universal thermodynamic principle.

In both classical and special relativistic kinetic theories, particles are typically required to remain ``on-shell,'' meaning that their mass-energy is fixed:
\eq{
{H}_{\eta} = 0 \iff p^0 = mc\gamma = mc\sqrt{1 + \frac{\bol{p}^2}{m^2c^2}}.
}

In a general relativistic context, however, it is natural to ask how collisions can be formulated when the underlying spacetime geometry itself influences the interaction process. 
Here, we propose to regard the geodesic Hamiltonian $\mc{H}$ (see equation~\eqref{gH}) as the physical energy and to incorporate interactions directly into its expression. 
In this picture, particles may temporarily deviate from geodesic motion during collisions, as their trajectories are influenced by non-gravitational forces. 
A detailed account of this formulation will be presented in Sec.~3.


\subsection{Liouville measure, entropy, and thermodynamics}
A central difficulty in relativistic thermodynamics is that the standard 
Liouville measure $dxdp$ is invariant under proper-time evolution,  but the reduced phase space measure $d\bol{x}dp$ is not invariant  
under coordinate-time evolution.  
As a consequence, the classical Shannon entropy $-\!\int f\log f\,d\bol{x}\,dp$ is not 
meaningful on $t=\mathrm{const}$ hypersurfaces.  
To formulate a collision operator that increases entropy with respect to the 
thermodynamic arrow of time~$t$, one must identify an invariant measure 
$J\,d\bol{x}dp$ adapted to the coordinate-time flow.  
This is the topic examined in this section.

Within the realm of Newtonian mechanics, there is a precise correspondence between kinetic theory and statistical mechanics on one hand, and thermodynamics on the other. 
This connection is built around the physical parameters (such as temperature) that characterize statistical distributions at thermodynamic equilibrium. 
However, kinetic theory may also be regarded as a standalone ``physical'' theory, 
one that does not require thermodynamics to be self-consistent or in agreement with experimental observations. 
Indeed, it is founded on microscopic particle dynamics and a specific set of statistical assumptions 
(e.g., the closure hypotheses that determine the form of the collision operator). 
Mathematically, kinetic theory describes the evolution of the distribution function according to the {coadjoint action} associated with the {Lie--Poisson structure} of the phase space (see, e.g., the discussion in \cite{Yos22}).

The possibility of formulating kinetic theory as an autonomous theory will be particularly helpful in this study, 
because general relativistic thermodynamics presents the fundamental difficulty that 
the first and second laws of thermodynamics---which describe how energy and entropy change in a system---imply the existence of a \ti{thermodynamic arrow of time} with respect to which such changes are defined.  
Moreover, the notions of energy and entropy themselves 
lose their classical meaning in a general relativistic context, 
preventing a straightforward and physically meaningful generalization of the thermodynamic laws.
Indeed, we remark that for a thermodynamics-type theory to hold, 
it is necessary to identify notions of macroscopic energy and entropy, where the latter represents an information measure encoding the (lack of) information about the internal state of the system~\cite{Shannon,Jaynes}. 
For example, black hole thermodynamics rests on the identification of the black hole surface area as the relevant entropy measure that cannot decrease~\cite{Bekenstein}. 
The core reason why a straightforward generalization of thermodynamics is not available is that 
general relativity lacks the coordinate-time-independent phase-space volume element (Liouville measure) that, for the reasons outlined below, would be needed to introduce a conventional notion of entropy $S\!\lrs{f}$ 
in terms of the Shannon information encoded in the distribution function $f\!\lr{x,p}$. 
Let us elaborate this point in detail.

First, recall that  
given a probability $p_i$ for the $i$th possible outcome of a random process,  
Shannon's information measure $S\lrs{p_i}$ is the unique functional satisfying  
continuity, monotonicity, and additivity,  
\eq{
S_{\rm Sh} = -\sum_i p_i \log p_i.
}
When dealing with continuous distributions $f$, the \ti{differential entropy} $S\lrs{f}$ is defined as  
\eq{
S = -\int f \log f\, dz, \label{dS}
}
where the integral is performed over the whole phase space $T^{\ast}M$.  
Here, the underlying assumption is that the phase space volume element  
\eq{
dz = dx^0 dx^1 dx^2 dx^3 \, dp_0 dp_1 dp_2 dp_3 = dx\,dp,
}
defines an \ti{invariant measure}:  
\eq{
\mf{L}_{\dot{z}} dz = \frac{\p \dot{z}^A}{\p z^A} \, dz = 0, \label{IM}
}
where $\mf{L}$ is the Lie derivative,  
\eq{
\dot{z} = \dot{z}^A \p_{z^A} = \frac{d z^A}{d\tau} \p_{z^A},
}
and $\p_{z^A}$ denotes the tangent vector in the $z^A$-direction.  
Note that if $dz$ were not an invariant measure, then differential entropy would not be well-defined,  
because the argument of the logarithm could not be identified with the infinitesimal probability $dP = f\,dz$.  
The canonical Hamiltonian structure of the equations of motion  
\eq{
\dot{z}^A = J_c^{AB} \frac{\p \mc{H}}{\p z^B} 
\iff 
\dot{x}^{\mu} = \frac{\p \mc{H}}{\p p_{\mu}}, 
~~~~
\dot{p}_{\mu} = -\frac{\p \mc{H}}{\p x^{\mu}},
}
ensures that condition \eqref{IM} is always satisfied (Liouville theorem).  
However, if we consider a general relativistic setting where $\p f / \p\tau = 0$,  
the differential entropy \eqref{dS} becomes trivial:  
for any isolated system with spacetime distribution function $f(x, p)$,  
the second law of thermodynamics becomes  
\eq{
\frac{dS}{d\tau} = 0.
}
Therefore, the quantity $S$ cannot be used to measure the uncertainty of the distribution function $f(x, p)$,  
unless one is willing to consider proper-time-dependent configurations $f(x, p, \tau)$—a possibility that is not considered in this work.

Now consider the perspective of an observer in a coordinate system $x=\lr{x^0,x^1,x^2,x^3}$ 
that measures the evolution of the distribution function $f\lr{x,p}$ with respect to coordinate-time $t = x^0 / c$, 
which now defines the thermodynamic arrow of time. 
Then, the phase space volume element $dz$ can no longer serve as an invariant measure, because it includes $dx^0$. 
The non-triviality of defining differential entropy for the reduced 7-dimensional phase space
\eq{
dV = dx^1 dx^2 dx^3\, dp_0 dp_1 dp_2 dp_3 = d\bol{x}dp
}
lies in finding an invariant measure. Specifically, an invariant measure 
\( J(\bol{x}, p) \, d\bol{x} dp \) that is independent of \( t \) 
is required. 
Indeed, differential entropy $S_t\lrs{f}$ could be defined as
\eq{
S_t = -\int f \log \left( \frac{f}{J} \right) \, d\bol{x} dp.\label{St}
}
The function \( J \) must satisfy a time-independent condition, $\p J/\p t=0$, to ensure that the argument $f/J$ inside the logarithm 
represents a pure probability (up to a constant factor related to the invariant volume element) 
encoding all the statistical information independently from the volume element $J\,d\bol{x}dp$.

On the other hand, for \( J\,d\bol{x}dp \) to be invariant, the function $J$ must satisfy the  condition:
\eq{
\frac{\p J}{\p t}+
\frac{\partial}{\partial x^{i}} \left( \frac{\dot{x}^{i}}{\gamma} J \right) 
+ \frac{\partial}{\partial p^{\mu}} \left( \frac{\dot{p}_{\mu}}{\gamma} J \right) 
= \lrc{\frac{J}{\gamma},\mc{H}}=0. \label{IMc}
}
Note that here a factor $\gamma$ appears because the time parameter for the equations of motion is given by $t$, 
\sys{
&\frac{d{x}^i}{dt} = \frac{1}{\gamma} \frac{\p \mc{H}}{\p{p}_i},\\
&\frac{d{p}_{\mu}}{dt} = -\frac{1}{\gamma} \frac{\p \mc{H}}{\p{x}^{\mu}},
}{EoM1}
However, in general, the solution \( J(t,\bol{x},p) \) of \eqref{IMc} will satisfy \( \partial J / \partial t \neq 0 \), 
because \( \partial \mc{H} / \partial t \neq 0 \). 

Now consider the collisionless Liouville equation in time $t$,
\eq{
\frac{\p}{\p x^{\mu}} \left( \frac{\dot{x}^{\mu}}{\gamma} f \right)
+ \frac{\p}{\p p_{\mu}} \left( \frac{\dot{p}_{\mu}}{\gamma} f \right) = \lrc{\frac{f}{\gamma},\mc{H}}=0. \label{Liouv}
}
After integration by parts and discarding boundary terms, we find that
\eq{
\frac{dS_t}{dt} =&
-\int \frac{\p f}{\p t} \left[ 1 + \log \lr{\frac{f}{J}}\right] \, d\bol{x} dp
+ \int f \frac{\partial \log J}{\partial t} \, d\bol{x} dp \\
=&
 \int \left[ \frac{\p}{\p x^i} \left( \frac{\dot{x}^i}{\gamma} f \right) 
+ \frac{\p}{\p p_{\mu}} \left( \frac{\dot{p}_{\mu}}{\gamma} f \right) \right] 
\left[ 1 + \log \lr{\frac{f}{J}}
\right] \, d\bol{x} dp  + \int f \frac{\p \log J}{\p t} \, d\bol{x} dp \\
=&\int \frac{f}{J}\lrs{
\frac{\p}{\p x^{\mu}}\lr{\frac{\dot{x}^{\mu}}{\gamma}J}+\frac{\p}{\p p^{\mu}}\lr{\frac{\dot{p}_{\mu}}{\gamma}J}}\,d\bol{x}dp\\
=&\int\frac{f}{J}\lrc{\frac{J}{\gamma},\mc{H}}\,d\bol{x}dp.
}
If $\p J/\p t=0$ (coordinate-time independent invariant measure), this expression 
does not vanish, despite the absence of collisions, because $\p\mc{H}/\p t\neq 0$ 
in general. This example highlights, through a kinetic argument, why the second law of thermodynamics 
does not manifest in the same form as in classical mechanics, due to the relativistic nature of time. 
This issue is also the reason why, e.g., in Synge theory~\cite{Israel}, 
the entropy law (H-theorem) arises as the non-negativity of the divergence of the entropy flux vector, 
$\nabla_{\mu} S^{\mu} \geq 0$, 
a form that deviates from the classical statement of the second law of thermodynamics. 

In this work, we resolve the entropy issue by observing that, if one foregoes the usual information–theoretic interpretation and allows a coordinate-time dependent reduced phase space invariant measure, one can define an entropy functional that is conserved in the collisionless case and increases in the presence of collisions—the desired thermodynamically consistent behavior. 
Indeed, setting $J=\gamma$ in \eqref{IMc} yields a time-dependent invariant measure $\gamma\,d\bol{x}dp$ and a corresponding entropy measure $S_t$ whose rate of change under the collisionless Liouville equation \eqref{Liouv} vanishes identically,
\eq{
\frac{dS_t}{dt}
=\int 
\frac{f}{\gamma}\lrc{1,\mc{H}}\,d\bol{x}dp
=0.
}
Remarkably, the condition for the existence of a coordinate-time independent invariant measure,
\eq{
\frac{\p \gamma}{\p t}
=\frac{\p^2 \mc{H}}{\p x^0\,\p p_0}
=\frac{1}{m}\frac{\p g^{0\mu}}{\p x^0}p_{\mu}+\frac{\p^2\Phi}{\p x^0\,\p p_0}=0,
}
is satisfied whenever $g$ admits a coordinate-time Killing symmetry and, in addition, the average interaction potential obeys $\p^2\Phi/\p x^0\,\p p_0=0$. 
Thus, the standard information–theoretic framework is recovered when the system develops a coordinate-time Killing symmetry.


We conclude this section with some remarks on conservation of total energy and the first law of thermodynamics. 
As for entropy, whenever $\p f/\p\tau=0$ and $\p\mc{H}/\p \tau=0$, a trivial notion of total energy \( U \) for the system can be obtained via the weighted sum of the Hamiltonians $\mc{H}$,
\eq{
U = \int f \mc{H} \, dz,
}
where, as usual, the integral is carried over the whole phase space. 
The first law of thermodynamics holds trivially:
\eq{
\frac{dU}{d\tau} = 0.
}
However, when treating \( t \) as the thermodynamic arrow of time, the total energy becomes
\eq{
U_t = \int f \mc{H} \, d\bol{x} dp,\label{Ut}
}
where the integral is carried over the $t$-level set \( T^*M_t = \lrc{z \in T^{\ast}M : x^0 = ct} \).  
For the collisionless Liouville equation \eqref{Liouv}, discarding boundary integrals, we find 
\eq{
\frac{dU_t}{dt} 
&= \int \left( \frac{\p f}{\p t} \mc{H} + f \frac{\p \mc{H}}{\p t} \right) \, d\bol{x} dp \\
&= \int \left\{ \left[ -\frac{\p}{\p x^i} \left( \frac{\dot{x}^i}{\gamma} f \right) - \frac{\p}{\p p_{\mu}} \left( \frac{\dot{p}_{\mu}}{\gamma} f \right) \right] \mc{H} + f \frac{\p \mc{H}}{\p t} \right\} \, d\bol{x} dp \\
&= \int f \left( \frac{\dot{p}_0}{\gamma} \frac{\p \mc{H}}{\p p_0} + \frac{\p \mc{H}}{\p t} \right) \, d\bol{x} dp = 0.
}
We will see in section~5 that $U_t$ remains an invariant of the Landau--Einstein system \eqref{LE}. 
Hence, a notion of conserved total energy can be defined at each coordinate-time instant  
by recognizing the Hamiltonian $\mc{H}$ as the physical energy of a particle.

The considerations above show that, in general relativity, the correct 
thermodynamic description is tied to the coordinate-time flow, to the 
invariant measure $\gamma\,d\bol{x}dp$, and to the use of the  
Hamiltonian $\mc{H}$ as the physically meaningful energy.  
These structural features determine both the form of the collision operator 
constructed in Section~4 and the structure of thermodynamic equilibria 
characterized in Section~6.


\section{Binary interactions in general relativistic spacetime}
The aim of this section is to explain how binary collisions are modeled in the present work.

First, recall that in the absence of non-gravitational forces, a massive particle with mass $m$ follows the geodesics associated with the metric tensor $g_{\mu\nu}$. 
The geodesic equations are generated by the geodesic Hamiltonian
\eq{
H=\frac{1}{2}mc^2+\frac{1}{2m}g^{\mu\nu}p_{\mu}p_{\nu},\label{H1}
}
according to
\sys{
&\dot{x}^{\mu}=\frac{dx^{\mu}}{d\tau}=\frac{\p H}{\p p_{\mu}}=\frac{1}{m}g^{\mu\nu}p_{\nu},\\
&\dot{p}_{\mu}=\frac{dp_{\mu}}{d\tau}=-\frac{\p H}{\p x^{\mu}}=-\frac{1}{2m}\frac{\p g^{\lambda\nu}}{\p x^{\mu}}p_{\lambda}p_{\nu}.
}{geq}
As usual, the dot symbol indicates differentiation with respect to the proper time $\tau$, which is defined according to the speed of light constraint 
\eq{
-c^2d\tau^2=g_{\mu\nu}dx^{\mu}dx^{\nu}.\label{pt}
}
The equivalence between the Hamiltonian system \eqref{geq} and the usual form of the geodesic equations can be verified by noting that $\dot{x}^{\mu}=p^{\mu}/m$ and
\eq{
\dot{p}^{\mu}=&\frac{d}{d\tau}\lr{g^{\mu\nu}p_{\nu}}=
\frac{p^{\alpha}p^{\beta}}{m}\lr{g_{\nu\beta}\frac{\p g^{\mu\nu}}{\p x^{\alpha}}-\frac{1}{2}g^{\mu\nu}\frac{\p g^{\lambda \eta}}{\p x^{\nu}}g_{\lambda\alpha}g_{\eta\beta}}\\=&\frac{p^{\alpha}p^{\beta}}{m}\lrs{\frac{1}{2}g^{\mu\nu}\lr{\frac{\p g_{\alpha\nu}}{\p x^{\beta}}-\frac{\p g_{\beta\nu}}{\p x^{\alpha}}}-\Gamma^{\mu}_{\alpha\beta}}=-\frac{p^{\alpha}p^{\beta}}{m}\Gamma_{\alpha\beta}^{\mu}, 
}
where $\Gamma_{\alpha\beta}^{\mu}$ denote the Christoffel symbols. 
Next, we note that the Lorentz factor is
\eq{
\gamma=\frac{dt}{d\tau}=\frac{1}{c}\frac{\p H}{\p p_0}=\frac{1}{mc}
g^{0\nu}p_{\nu},
}
and reformulate the geodesic equations \eqref{geq} as a system of ODEs with respect to coordinate-time $t$, 
\sys{
&\frac{dx^{\mu}}{dt}=\frac{1}{m\gamma}g^{\mu\nu}p_{\nu},\\
&\frac{dp_{\mu}}{dt}=-\frac{1}{2m\gamma}\frac{\p g^{\lambda\nu}}{\p x^{\mu}}p_{\lambda}p_{\nu}.
}{geqt}
Observe that the Hamiltonian $H$ given in equation \eqref{H1} remains an invariant of system \eqref{geqt}, because
\eq{
\frac{dH}{dt}=\frac{1}{\gamma}\frac{dH}{d\tau}=0.
}

Now consider two particles of mass \( m \) with phase space coordinates \( (x, p) \) and \( (x', p') \). We assume that these particles interact via a scalar interaction potential energy
\begin{equation}
V(x, x', p, p'),
\end{equation}
which respects causality, meaning that the support of \( V \) lies within the union of the past and  future light cones of \( x \). 
This condition
ensures that the interaction propagates at a finite speed not exceeding the speed of light \( c \).

As an example, if the two particles carry electric charge \( q \), a possible covariant generalization of the Coulomb potential energy is
\eq{
V(x,x',p,p') =
\big[\Theta(x \succ x') + \Theta(x' \succ x)\big]\,
A_{\varepsilon}\,
\frac{q^{2}}{\sqrt{2\abs{\sigma(x,x')} + \varepsilon^{2}}}.
\label{VCg}
}
where:
\begin{itemize}
  \item $\sigma(x,x')$ is Synge's world function, equal to one-half the squared geodesic distance between $x$ and $x'$,
  \item $\Theta(x \succ x')$ is a generalized step function, equal to $1$ if $x$ lies in the causal future of $x'$ and $0$ otherwise,
  \item $\varepsilon>0$ is a small regularization parameter,
  \item $A_{\varepsilon}$ is an $\varepsilon$-dependent normalization constant chosen such that the potential
        satisfies the appropriate Maxwell equation (e.g.\ reproducing a delta-function source in the
        $\varepsilon\to 0$ limit).
\end{itemize} 

This potential depends only on the Synge world function and enforces causality through the
step functions by excluding spacelike or acausal pairs $(x,x')$.  In the Minkowski limit
and for equal coordinate-times, \eqref{VCg} reduces to a regularized
Coulomb interaction $q^{2}/|\boldsymbol{x}-\boldsymbol{x}'|$.  
Notice the symmetry $V\lr{z,z'}=V\lr{z',z}$, which expresses the fact that the potential energy remains the same if the identical particles exchange their phase space position.  


We emphasize that the precise form of \( V \) is not essential for the developments that follow, provided that $V$ is symmetric under exchange of particle positions,
\eq{
V\lr{z,z'}=V\lr{z',z}.\label{Vsym}
}
and it is \ti{speed of light compatible}, as defined below.  
\begin{mydef}
A Hamiltonian $\mc{H}\lr{x,p}$ is speed of light compatible whenever
\eq{
\lrc{\chi,\mc{H}}=0,~~~~\chi=c^2+g_{\mu\nu}\frac{\p \mc{H}}{\p p_{\mu}}\frac{\p\mc{H}}{\p p_{\nu}}.\label{slc}
}
If a speed of light compatible Hamiltonian $\mc{H}$ is derived from the geodesic Hamiltonian $H$ by including the effect of an interaction potential energy $V$, 
we say that $V$ is speed of light compatible.
\end{mydef}
Thanks to its general structure, the potential $V$ can also accommodate magnetic or velocity-dependent interactions. 
The key idea is that \( V \) represents a fundamental, manifestly covariant interaction energy, which when added to the geodesic Hamiltonians yields the total relativistic Hamiltonian of the system.

Now observe that, on average, the interaction potential energy felt by a particle is given by the integral~\eqref{Phi}, which may also be written as
\eq{
\Phi\lr{x,p} = \int f\lr{x',p'}\, V\lr{x,x',p,p'}\, dx'\, dp',
}
where \( f \) is the distribution function of an ensemble of \( N \) particles to which the particle itself belongs. As previously explained, because the support of the function $V$ is limited to causally related events, 
the averaged potential energy \( \Phi \) inherits the causality properties of \( V \). The average total energy of the particle is therefore given by the single particle Hamiltonian \( \mc{H} \) defined in equation~\eqref{gH},
\eq{
\mc{H} = H+\Phi=\frac{1}{2} mc^2 \left( 1 + \frac{g^{\mu\nu}p_{\mu}p_{\nu}}{m^2c^2} \right) + \Phi.
}
Here, it is crucial to emphasize that Hamilton's equations for the Hamiltonian $\mc{H}$, 
\eq{
&\dot{x}^{\mu}=\frac{p^{\mu}}{m}+\frac{\p\Phi}{\p p_{\mu}},\\
&\dot{p}_{\mu}=-\frac{1}{2m}\frac{\p g^{\lambda\nu}}{\p x^{\mu}}p_{\lambda}
p_{\nu}-\frac{\p\Phi}{\p x^{\mu}},}
correctly reproduce the particle equations of motion in both the Minkowski and classical limits under appropriate choice of $\Phi$. 

Let \( f_{N-2} \) denote the distribution function of the statistical ensemble comprising \( N-2 \) particles. Because we assume \( N \gg 2 \), we may take \( f_{N-2} \approx f \).  Recalling that $z=\lr{x,p}$, we may then express the total energy of two interacting particles as
\eq{
&
{H}(z) + \int f_{N-2}(z')\, V(z,z')\, dz' + V(z,z') + \int f_{N-2}(z)\, V(z',z)\, dz + {H}(z') \\
&\approx {H}(z) + \Phi(z) + V(z,z') + \Phi(z') + {H}(z').
}
We thus define the pair Hamiltonian
\eq{
\mc{H}_{12} = \mc{H}+ V + \mc{H}', \label{pH}
}
where  
the prime indicates evaluation at \( z' \).

Evidently, the equations of motion of the two charged particles with respect to coordinate-time $t$, and  including the binary interaction, can be obtained as
\eq{
\frac{dz^A}{dt} =\frac{1}{\gamma} J_c^{AB} \frac{\p  \mc{H}_{12}}{\p z^{B}}, \qquad
\frac{d{z}'^{A}}{dt} = \frac{1}{\gamma'}J_c^{AB} \frac{\p\mc{H}_{12}}{\p z'^B}.
}
Naturally, the interacting particles must still  satisfy the speed-of-light constraint expressed in Eqs. \eqref{pt}. 
This imposes restrictions on the admissible form of $V$, which must be speed of light compatible  \eqref{slc}, reflecting the physical requirement that interactions cannot violate the bound set by $c$. 
Explicitly, Eq.~\eqref{slc} yields
\eq{
\lrc{g_{\mu\nu}\lr{\frac{p^{\mu}}{m}+\frac{\partial \Phi}{\partial p_{\mu}}}
\lr{\frac{p^{\nu}}{m}+\frac{\partial\Phi}{\partial p_{\nu}}},\frac{1}{2m}g^{\mu\nu}p_{\mu}p_{\nu}+\Phi}=0,\label{pt2}
}
which constrains the possible interaction potentials $V$ through $\Phi$. 
During binary interactions, $\Phi$ should be replaced by $\Phi+V$. However, the constraint involving $V$ is not explicitly needed for the construction of the collision operator.

In many familiar settings, the speed-of-light constraint $\chi=0$ is encoded by the conservation of a Hamiltonian
“mass–shell” function $\mc{H}$. For instance, if the average interaction is encoded in a four-potential $A_\mu$
(as in electromagnetism), then
\eq{
\Phi=-\frac{q}{m}\,p_\nu A^\nu+\frac{q^2}{2m}\,A_\nu A^\nu
\;\;\Longleftrightarrow\;\;
\mc{H}=\frac{1}{2}m c^2+\frac{1}{2m}\,g^{\mu\nu}\!\big(p_\mu-qA_\mu\big)\!\big(p_\nu-qA_\nu\big),\label{EM}
}
with $q$ the electric charge. Hence,
\eq{
\chi=c^2+g_{\mu\nu}\dot{x}^{\mu}\dot{x}^{\nu}=\frac{2}{m}\mc{H},
} 
which automatically enforces speed of light compatibility $\lrc{\chi,\mc{H}}=0$. 
Nevertheless, the physical constraint needed in the present theory is the speed of light normalization $\chi=0$; we do not impose $\mc{H}=0$ and $\chi\neq 2\mc{H}/m$ in general.

In principle, the speed of light constraint $\chi=0$ can be solved for $p_0$; when such a solution exists, we denote it by 
$p_0^{m}\lr{x,\boldsymbol{p}}$. 

Finally, we 
observe that 

\eq{
\frac{d\chi}{dt}=\frac{1}{\gamma}\lrc{\chi,\mc{H}_{12}}=\frac{1}{\gamma}\frac{\p\chi}{\p z}\cdot J_c\cdot\frac{\p\mc{H}_{12}}{\p z}
=0, 
}
for any speed of light compatible Hamiltonian  $\mc{H}_{12}$. 
Here, $\dot{z}$ is orthogonal to $\chi_z=\p \chi/\p z$.
This fact suggests that the conservation of $\chi$ can be regarded as a property of relativistic phase space, 
or, in the language of noncanonical Hamiltonian mechanics, that the quantity $\chi$ effectively behaves as a Casimir invariant \cite{Morrison82,Littlejohn}. 
The  ``physics" of the system is self-contained on the
submanifold $\chi=0$.  



\section{A general relativistic collision operator}
The purpose of this section is to derive the Landau–Einstein collision operator \eqref{LEC}, under the assumption of \ti{grazing scatterings}, i.e., binary collisions resulting in small deviations of the particles' phase space trajectories.  This assumption is analogous to the hypothesis of grazing Coulomb collisions used in the derivation of the classical Landau operator. 
 Below, we follow, mutatis mutandis, the construction of~\cite{SatoPJM2} for the case of classical collisions in general noncanonical phase spaces. 
The key difference is that 
the only assumption on the symmetries  of the potential energy $V$ is \eqref{Vsym}. 

\subsection{Grazing scatterings}

As a first step, we evaluate the phase space displacement 
$\delta z = z'' - z$ 
induced by a grazing collision event that sends a particle initially at \(z\) to a new position \(z''\).
To proceed, we must first introduce the notion of grazing scattering in the present framework. 
We take \(t\) as the time parameter and
consider the orbits $z(t)$ and $z'(t)$ of two colliding particles at the same coordinate-time $t=t'$.  
Along the orbits \(z(t)\) and \(z'(t)\), 
we consider an observable quantity
\(\mc{I}\!\left(z(t), z'(t)\right)\),
and define

\begin{mydef}
An $\mc{I}$-grazing scattering satisfies
\begin{equation}
\frac{1}{\mc{I}_{-}}\int_{-\infty}^{+\infty}\frac{d\mc{I}}{dt}\,dt
=
O(\varepsilon),
\label{GS}
\end{equation}
where $\mc{I}_{-}=\lim_{t\to -\infty}\mc{I}(z(t),z'(t))$, and $\varepsilon>0$ is a small positive constant (ordering parameter).
We say that $\mc{I}$ is a grazing invariant. 
\end{mydef} 

Let $L^A\sim\mc{H}/\p_{z^A}\mc{H}$ denote the characteristic scale 
in the $z^A$ direction {($A=0,\dots,7$)} associated with the Hamiltonian $\mc{H}$, with $L^0=T$.  
Similarly, let $\ell^A\sim V/\p_{z^A}V$ denote the characteristic scale in the $z^A$ direction associated with the binary interaction energy $V$, and let the \ti{collision time} $\tau_c\sim\ell^0$ be the characteristic coordinate-time interval of a binary collision.  
We assume that binary interactions are localized, in the sense that
\begin{equation}
\frac{\mc{H}}{\p_{z^A}\mc{H}} \sim L^A \gg \frac{V}{\p_{z^A}V} \sim
\ell^A,\label{O1}
\end{equation}
and
\begin{equation}
\frac{\ell^A}{L^A} \sim \epsilon,\label{O2}
\end{equation}
with $\epsilon>0$ a small real constant.  
 We further require that $V$ decays rapidly to zero for 
 $\lr{z^A-z^A{}'}^2>\ell^{A2}$, and 
 observe 
 that 
the collision time $\tau_c$ is small compared to the 
time scale $T$, namely
\begin{equation}
\frac{\tau_c}{T} \sim \epsilon .\label{O3}
\end{equation} 
In this setting, the grazing scattering condition \eqref{GS} becomes
\begin{equation}
\frac{1}{\mc{I}_{t}}
\int_{t}^{t+\tau_c}\!\left(\frac{d\mc{I}}{dt}
\right)_{t=\mf{t}}d\mf{t}
=O(\epsilon^2),\label{GS2}
\end{equation}
where, for later convenience, we set $\varepsilon=\epsilon^2$ (so that the phase space displacement caused by a binary collision is a first order effect). Here $\mc{I}_{t}=\mc{I}(z(t),z'(t))$, with $t$ denoting the time at which the binary interaction begins. 
In the following, we simplify the notation by adopting vector calculus notation,
\begin{equation}
J_c\cdot\mc{I}_z=J_c \cdot \frac{\p \mc{I}}{\p z}
=J_cd\mc{I}= J_c^{AB}\frac{\p \mc{I}}{\p z^B}\,\p_{z^A},
\end{equation}
and so forth, where $\cdot$ denotes contraction.  
Furthermore, given a function $F\lr{t}$, we shall write 
\eq{
\int_{t}^{t+\tau_c}F\lr{\mf{t}}\,d\mf{t}=\int_{\tau_c}F\,dt
}

Let us now return to the phase space displacement $\delta z$ caused by a collision. 
A 
key observation is that, because the chosen time parameter is the coordinate-time $t$ (and not the proper time $\tau$), 
we must set $\delta x^0=0$ in $\delta z$. 
Indeed, for a coordinate-time observer the phase space displacement $\delta z$ 
describes the change in phase space position $\lr{\bol{x},p}$ during a collision event spanning a coordinate-time interval $\tau_c$. 
Including $\delta x^0$ would amount to double counting because, by our choice of time parametrization,  
$dx^0/dt=c$ holds with or without collisions. 
That is, the coordinate-time evolution term in 
$\lrc{f/\gamma,\mc{H}}$,
\eq{
\frac{\p f}{\p t}
=\frac{\p}{\p x^0}\lr{f
\frac{dx^0}{d t}
},
}
already accounts for the entire contribution to the coordinate-time component of the   
phase-space velocity $dx^0/dt=c$, and there is no
additional collisional flux in the $x^0$ direction.


The conservation of the speed of light constraint  $\chi=c^2+g_{\mu\nu}\mc{H}_{p_{\mu}}\mc{H}_{p_{\nu}}=0$ and the condition $\delta x^0=0$ can be enforced simultaneously by noting that, in this setting, 
the collision process can be implemented by replacing the canonical Poisson tensor $J_c$ with the projected antisymmetric tensor
\eq{
J_{\Sigma}=\mathbb{P}_{\Sigma}\cdot J_c\cdot\mathbb{P}_{\Sigma},\label{JSIGMA}
}
where we defined the projector 
\eq{
\mathbb{P}_{\Sigma}:T_z\!\left(T^{\ast}M\right)\rightarrow T_z\!\left(T^{\ast}M\right),\label{Proj}
}
onto the $6$-dimensional distribution  
\eq{
\Sigma_z=\bigl\{\delta z\in T_z\!\left(T^{\ast}M\right):d\chi(\delta z)=0,\; dx^0(\delta z)=0\bigr\},
}
according to
\eq{
\mathbb{P}_{\Sigma}=I-\p_0\otimes dx^0-\p_{\chi}\otimes d\chi,
}
where $I$ is the identity, and 
we used new coordinates $y=\lr{x,\chi,\bol{p}}$. In components,
\eq{
\lr{\mathbb{P}_{\Sigma}}^{A}_{~B}&=\delta^A_{~B}-\delta^A_{~0}\delta^0_{~B}-\delta^A_{~\chi}\delta^{\chi}_{~B},\\
J_{\Sigma}^{AD}&=J_c^{AD}-J^{A0}_c\delta^D_{~0}-J^{A\chi}_c\delta^D_{~\chi}-J_c^{0D}\delta^A_{~0}-J^{\chi D}_c\delta^A_{~\chi}+J^{0\chi}_c\lr{\delta^A_{~0}\delta^D_{~\chi}-\delta^A_{~\chi}\delta^D_{~0}}.\label{components}
}
Notice the properties $\mathbb{P}^2_{\Sigma}=\mathbb{P}_{\Sigma}$,  
$J_{\Sigma}{dx^0}=J_{\Sigma}{d\chi}=0$, as well as \eq{J_{\Sigma}dV=J_{\Sigma}\mathbb{P}_{\Sigma}dV=J_{\Sigma}\cdot\mathbb{P}_{\Sigma}\cdot \frac{\p V}{\p z}=J_{\Sigma}^{AD}\lr{\mathbb{P}_{\Sigma}}^{C}_{~D}\frac{\p V}{\p y^C}\p_{y^A}.} 
Taking physical units such that $T\sim\ell^A\sim 1$ and $\tau_c\sim 1/L^A\sim\epsilon$ for $A\neq 0$, 
it follows that
\eq{
\delta z=\int_{\tau_c}\frac{1}{\gamma}
J_{\Sigma}
\cdot \frac{\p\mc{H}_{12}}{\p z}\,dt=
\frac{1}{\gamma_t}
J_{\Sigma t}\cdot\int_{\tau_c}\mathbb{P}_{\Sigma}\cdot\frac{\p V}{\p z}\,dt+O\lr{\epsilon^2}, \label{dz00}
}
where, in the last passage, we used the hypothesis that 
collisions are localized, giving $\gamma^{-1}J_{\Sigma}=\gamma_t^{-1}J_{\Sigma t}+O\lr{\epsilon}$, and where, as usual, the lower index indicates evaluation at time $t$. 
We observe that the term $\p\mc{H}_{12}/\p x^0$ does not contribute to $\delta z$ due to the presence of the projector $\mathbb{P}_{\Sigma}$.  
Now recall that, by hypothesis, $\p V/\p z^A\sim V/\ell^A$ while $\p\mc{H}/\p z^A\sim \mc{H}/L^A$. 
Hence,
omitting the lower index $t$ to simplify the notation,  
the phase space displacement \eqref{dz00} caused by a collision event can be written as 
\eq{
\delta z
=\frac{1}{\gamma}J_{\Sigma}\cdot\int_{\tau_c}V_z
\,dt+O\lr{\epsilon^2}
, \label{dz0}
}
where we defined 
\eq{
V_z = \mathbb{P}_{\Sigma} \cdot \frac{\partial V}{\partial z}=\lr{\mathbb{P}_{\Sigma}}^C_{~D}\frac{\p V}{\p y^C}dy^D.
}
We emphasize that this notation is introduced solely to simplify the appearance of expressions, and that the projector $\mathbb{P}_{\Sigma}$ is implicitly present. This differs from the general notation $f_z = \partial f / \partial z$, which will be used for all quantities $f$  other than $V$.

We now assume that the collision process is $\mc{I}$-grazing for some additive quantity $\mc{I}=\mc{A}+\mc{A}'$, where $\mc{A}\lr{z}$ will be specified later. Introducing the vector field
\eq{
\xi=\frac{1}{\gamma}J_{\Sigma}\cdot\mc{A}_z,\label{xi}
}
with $\mc{A}_z=\p \mc{A}/\p z$, 
we have
\eq{
\frac{d\mc{I}}{dt}=\frac{1}{\gamma}\mc{A}_z\cdot J_{\Sigma}\cdot\frac{\p\mc{H}_{12}}{\p z}+\frac{1}{\gamma'}\mc{A}_z'\cdot J_{\Sigma}'\cdot\frac{\p\mc{H}_{12}}{\p z'}=-\xi\cdot V_z-\xi'\cdot V_z'+O\lr{\epsilon},
}
where we used the symmetry $V=V'$ (eq. \eqref{Vsym}), which implies  
\eq{
V_z'=\mathbb{P}_{\Sigma}'\cdot\lr{\frac{\p V}{\p z}}'=\mathbb{P}_{\Sigma}'\cdot\frac{\p V}{\p z'}. } 
On the other hand, the $\mc{I}$-grazing scattering assumption implies that
\eq{
O\lr{\epsilon^2}=\int_{\tau_c}\frac{d\mc{I}}{dt}\,dt=-\xi_t\cdot \int_{\tau_c}V_z\, dt-\xi_t'\cdot\int_{\tau_c}V_z'\,dt+O\lr{\epsilon^2}.
}
Omitting the lower index $t$ again, we have thus found that
\eq{
\xi\cdot\int_{\tau_c}V_z\,dt=-
\xi'\cdot\int_{\tau_c}V_z'\,dt+O\lr{\epsilon^2}. 
}
This last expression implies that, up to $O\lr{\epsilon^2}$ terms, we may express the component of $\int_{\tau_c} V_z\,dt$ along $\xi$ in terms of $\xi'\cdot\int_{\tau_c} V_z'\,dt$ and vicecersa.
Hence, if we introduce the vector field
\eq{
q=
{\xi}-{\xi'}=\lr{\xi^A-\xi^A{}'}\p_{z^A},\label{q}
}
and the orthogonal projector $\mathbb{P}_q^{\perp}$ as
\eq{
\lr{\mathbb{P}_q^{\perp}}_A{}^B=\delta_A{}^B-\frac{q_Aq^B}{\abs{q}^2},} 
where $q_A=\delta_{AB}q^B$,
$\abs{q}^2=q_Aq^A$, $q^{\flat}=q^B\delta_{BA}d{z^A}$, 
we may write
\eq{
\int_{\tau_c}V_z\,dt=\mathbb{P}_q^{\perp}\cdot \int_{\tau_c}V_z\,dt+\frac{q\cdot\int_{\tau_c}V_z\,dt}{\abs{q}^2}q^{\flat}=
\mathbb{P}_q^{\perp}\cdot \int_{\tau_c}V_z\,dt-\frac{\Delta \cdot\xi'}{\abs{q}^2}q^{\flat}+O\lr{\epsilon^2},
}
where
\eq{
\Delta =\int_{\tau_c}V_z\,dt+\int_{\tau_c}V_z'\,dt.\label{delta}
}
At leading order, the phase space displacement \eqref{dz0} caused by a binary collision event can thus be written as 
\begin{equation}
\delta z
=\frac{1}{\gamma}J_{\Sigma}\cdot\mathbb{P}\cdot\int_{\tau_c} V_z\,dt, 
\label{dz}
\end{equation}
where we have introduced the operator $\mathbb{P}$ such that 
\eq{
\mathbb{P}\cdot \int_{\tau_c}V_z\,dt=\mathbb{P}_q^{\perp}\cdot \int_{\tau_c}V_z\,dt-\frac{\Delta \cdot\xi'}{\abs{q}^2}q^{\flat}.\label{mathbbP}
}
Now notice that $\Delta=\Delta '$, $q^A=-q^A{}'$, 
and that $\lr{\mathbb{P}^{\perp}_q}_A^{~B}=\lr{\mathbb{P}^{\perp}_{q'}}_A^{~B}$. Hence, 
\eq{
\delta z'=\frac{1}{\gamma'}J_{\Sigma}'\cdot\mathbb{P}'\cdot\int_{\tau_c}V_z'\,dt,\label{dzp}
}
with 
\eq{
\mathbb{P}'\cdot\int_{\tau_c}V_z'\,dt=\mathbb{P}^{\perp}_q\cdot\int_{\tau_c}V_z'\,dt-\frac{\Delta\cdot\xi}{\abs{q}^2}q^{\flat}{}'.
}


We now note that the ordering assumptions \eqref{O1}, \eqref{O2}, and \eqref{O3}
imply that binary scatterings are $\mc{I}$-grazing, with $\mc{A}=\mc{H}$ and   $\mc{I}=\mc{H}+\mc{H}'$ the average total energy of the two colliding particles. Indeed, 
we have 
\begin{equation}
\frac{d\mc{I}}{dt} 
= \frac{1}{\gamma}\frac{\p \mc{H}}{\p z}\cdot J_{\Sigma} \cdot \frac{\p V}{\p z}+\frac{1}{\gamma'}\frac{\p\mc{H}'}{\p z'}\cdot{J}_{\Sigma}'\cdot\frac{\p V}{\p z'}
= -\xi \cdot 
V_z-\xi'\cdot 
V_{z}'
=O\lr{\epsilon}.
\end{equation} 
It follows that 
\begin{equation}
\frac{1}{\mc{I}_t}\int_{\tau_c}\frac{d\mc{I}}{dt}\,dt=-\frac{1}{\mc{I}_t}\,{\xi
}_{t} \cdot \int_{\tau_c}V_z\,dt-\frac{1}{\mc{I}_t}\xi'
_{t}\cdot\int_{\tau_c}V_z'\,dt+O\lr{\epsilon^3}
= O(\epsilon^2),
\end{equation}
which is the grazing scattering condition \eqref{GS2}. 
As it will be shown in 
Sec. 6, the additive quantity 
$\mc{I}=\mc{H}+\mc{H}'$ 
leads to the Landau--Einstein distribution \eqref{LEd}  at thermodynamic equilibrium.

If, on the other hand, $\mc{H}$ is replaced
by the special relativistic energy $p_0$ so that collisions are $\mc{I}$-grazing with $\mc{I}=p_0+p_0'$, one arrives at Landau--Synge type distributions \eqref{LSd} (see Sec. 6). However, we emphasize that, even if the ordering assumptions \eqref{O1}, \eqref{O2}, and \eqref{O3} are satisfied, they do no guarantee $p_\mu+p_\mu'$-grazing collisions because 
$dp_\mu/dt+dp_\mu'/dt=\gamma^{-1}p_{\mu z}\cdot J_{\Sigma}\cdot V_z+\gamma'{}^{-1}p_{\mu z'}'\cdot J_{\Sigma}'\cdot V_{z}'+O\lr{\epsilon}=O\lr{1}$, which does not allow the projection of the $1$-form $\int_{\tau_c} V_z\,dt$ in the expression of $\delta z$. 
We verify the claim for $p_0 + p_0'$ in the simplest case of flat spacetime,
$g=\eta$, with $\Phi=0$ and the mass--shell constraint
\eq{
\chi = c^2 + \frac{\eta^{\mu\nu}p_\mu p_\nu}{m^2}.
}
In differential--form notation, the contraction
$p_{0z}\!\cdot\! J_{\Sigma}\!\cdot\! V_z$ reads
\eq{
p_{0z}\!\cdot\! J_{\Sigma}\!\cdot\! V_z
=dp_0\lr{\mathbb{P}_{\Sigma}{J_c\mathbb{P}_{\Sigma}dV}}. 
}
By $g=\eta$, 
and using the component expression \eqref{components} 
in the coordinates $y=(x,\chi,\bol{p})$, we find
\eq{
dp_0&\lr{\mathbb{P}_{\Sigma}J_c\mathbb{P}_{\Sigma}dV}\\=&
\frac{\p p_0}{\p p_i}dp_i\lr{\lr{J_c^{AD}V_{y^D}-J_c^{A0}V_{x^0}-J_c^{A\chi}V_{\chi}-J_c^{0D}\delta^A_{~0}V_{y^D}-J_c^{\chi D}\delta_{~\chi}^AV_{y^D}+J_c^{0\chi}\lr{\delta^A_{~0}V_{\chi}-\delta^A_{~\chi}V_{x^0}}}\p_{y^A}}\\=&
\frac{\p p_0}{\p p_i}\lr{J_c^{p_i D}V_{y^D}-J_c^{p_i 0}V_{x^0}-J_c^{p_i\chi}V_{\chi}}.
}
Noting that 
\eq{
J_c=\p_{\mu}\w \p_{p_{\mu}}=\p_i\w\p_{p_i}+\chi_{p_i}\p_i\w \p_{\chi}+\chi_{p_0}\p_0\w \p_{\chi}, 
}
and 
using 
the relations 
\eq{
p_0 = -\sqrt{\,m^2(c^2 - \chi) + \boldsymbol{p}^2\,},
\qquad
dp_0 = \frac{\partial p_0}{\partial p_i}\, dp_i+\frac{\p p_0}{\p\chi}\,d\chi
      = \frac{p^i}{p_0}\, dp_i-\frac{m^2}{2p_0}\,d\chi,
}
we find
\eq{
dp_0\!\left(\mathbb{P}_{\Sigma}J_c\mathbb{P}_{\Sigma} dV\right)
= \frac{p^i}{p_0}J_c^{p_i i}V_{y^i}=-\frac{p^{i}V_{x^i}}{p_0}.
}
Thus $dp_0/dt+dp_0'/dt$ is a generally nonzero $O\lr{1}$ term. 
Note also that this result remains valid even if $V$ is required to be speed of light compatible. 
Indeed, $\{V,\chi\}=0$ implies $p_0V_{x^0} = p^iV_{x^i}$, leading to 
${dp_0}/{dt} + {dp_0'}/{dt} = 
-\,\gamma^{-1}V_{x^0} - \gamma'{}^{-1}V_{x^0}'=
O\lr{\epsilon^{-1}}$.


\subsection{Collision operator}
We shall now use the derived displacements \eqref{dz} and \eqref{dzp} to construct the desired collision operator.

In the absence of collisions, the equation governing the particle distribution function $f$ encodes conservation of particle number and can be written as
\begin{equation}
\left(\frac{\p}{\p t} + \mf{L}_{\gamma^{-1}\lr{\dot{x}^i\p_i+\dot{p}_{\mu}\p_{p_{\mu}}}}\right) 
f \, d\bol{x} dp
= \left[
\frac{\p f}{\p t}
+ \frac{\p}{\p x^i}\!\left(f\frac{dx^i}{dt}\right)
+ \frac{\p}{\p p_\mu}\!\left(f\frac{dp_\mu}{dt}\right)
\right]
d\bol{x} dp = 0,
\end{equation}
where $\mf{L}$ denotes the Lie derivative on $T^*M_t$.   
Since $dz/dt = \gamma^{-1}\, J_c \cdot \p_z \mc{H}$, this equation is equivalent to Liouville’s equation
\begin{equation}
\left\{ \frac{f}{\gamma}, \mc{H} \right\} = 0,
\end{equation}
with Poisson bracket $\lrc{\cdot,\cdot}$ given by  \eqref{cPB}, and where it is understood that the Lorentz factor is the one determined by $\mc{H}$ as $\gamma=\p\mc{H}/\p p_0$.  
Once collisions are taken into account, the governing equation for $f$ takes the form
\begin{equation}
\left\{ \frac{f}{\gamma}, \mc{H} \right\} 
= \mathscr{C}(f,f),
\end{equation}
where $\mathscr{C}(f,f)$ denotes the collision operator.

Now consider two particles with phase space coordinates $z$ and $z'$ undergoing a collision that results 
in new positions $z'' = z + \delta z$ and $z''' = z' + \delta z'$, respectively. 
Since $t$ is chosen as the time parameter, 
the invariant measure $\gamma \,d\bol{x}dp$ acquires a factor $\gamma$. 
For the subsequent calculations it is convenient to introduce the 
distribution function 
\eq{
\mc{F} = \frac{f}{\gamma},
}
which is defined on the invariant measure $\gamma\,d\bol{x}dp$.
The collision operator for such a process can be written in the general form
\begin{equation}
\mathscr{C}(f,f) = \int 
\Big[ \mc{V}(z'',z''';z,z')\, \mc{F}(z'')\mc{F}(z''') 
     - \mc{V}(z,z';z'',z''')\, \mc{F}(z)\mc{F}(z') \Big]\,
dz'\,dz''\,dz''',
\label{Cop0}
\end{equation}
where $\mc{V}(z,z';z'',z''')$ is the so–called \emph{scattering volume density per unit (proper) time}. 
The quantity $\mc{V}(z'',z''';z,z')\,dz'\,dz''\,dz'''$ represents the phase space volume, per unit time, accessible to a particle initially at $z$ for scatterings with a target at $z'$ that produce the outcomes $z''$ and $z'''$, respectively. 
In particular, we remark that $\mc{F}\mc{V} dz'dz''dz'''$ has dimensions of $\tau^{-1}$. 
As usual, the first term in \eqref{Cop0} describes the \emph{inflow} of particles into the state $(z,z')$, while the second term represents the \emph{outflow} of particles leaving $(z,z')$. 

Because the equations governing binary interactions are time–reversible, reversing the time parameter $t \to -t$ simply inverts the order of collisions.  
Hence the scattering volumes are unchanged, and we may assume
\eq{
\mc{V}(z'',z''';z,z')\,dz'\,dz''\,dz''' 
= \mc{V}(z,z';z'',z''')\,dz'\,dz''\,dz'''.
}
With this symmetry, the collision operator reduces to
\begin{equation}
\mathscr{C}(f,f) 
= \int \mc{V}(z,z';z'',z''') 
   \Big[ \mc{F}(z'')\mc{F}(z''') - \mc{F}(z)\mc{F}(z') \Big]\,
   dz'\,dz''\,dz''' .
\label{Cop1}
\end{equation}
 Notice that the integrals here also involve integration with respect to the coordinate-time.  
This reflects the fact that interactions are not instantaneous: $\mc{V}$ does not vanish as long as $z$ and $z'$ are causally related.  
By contrast, for classical {(instantaneous) collisions} one has
\eq{
\mc{V} \;\propto\; 
\delta(x^0 - x^0{'})\,\delta(x^0 - x^0{''})\,\delta(x^0 - x^0{'''}) .
}
In particular, the Boltzmann collision operator can be obtained by setting
\eq{
\mc{V}=&m^{-6}{\sigma}_{\rm B}\lr{\bol{v},\bol{v}';\bol{v}'',\bol{v}'''}\abs{\bol{v}-\bol{v}'}\\
&\delta(x^0 - x^{0}{'})\,\delta(x^0 - x^{0}{''})\,\delta(x^0 - x^{0}{'''})
\delta\lr{\bol{x}-\bol{x}'}\delta\lr{\bol{x}''-\bol{x}'''}\delta\lr{\bol{x}-\bol{x}''}\\
&\delta\lr{p_0'+mc}\delta\lr{p_0''+mc}\delta\lr{p_0'''+mc}, 
}
where $\sigma_{\rm B}$ is the scattering cross section and $\bol{v}=\bol{p}/m$. 

Introducing the notation $\mc{F}_{zz}=\p^2 \mc{F}/\p z\p z$ and expanding the integrand in   equation \eqref{Cop1} to second order in $\delta z$ and $\delta z'$ we obtain
\eq{
\mathscr{C}\lr{f,f}=\int\mc{V}\lr{\mc{F}\mc{F}_z'\cdot\delta z'+\mc{F}'\mc{F}_z\cdot\delta z+\frac{1}{2}\mc{F}\delta z'\cdot \mc{F}_{zz}'\cdot\delta z'+\mc{F}_z\cdot\delta z \,\mc{F}_z'\cdot\delta z'+\frac{1}{2}\mc{F}'\delta z\cdot \mc{F}_{zz}\cdot\delta z}\,dz'dz''dz'''.
}
After some manipulations, it can be shown that this expression is equivalent to
\eq{
\mathscr{C}\lr{f,f}=&
\int\mc{V}\lr{\mc{F}\mc{F}_z'\cdot\delta z'+\mc{F}'\mc{F}_z\cdot\delta z}\,dz'dz''dz'''\\
&-\frac{1}{2}\int \mc{F}\mc{F}_z'\cdot\lrs{\frac{\p}{\p z'}\cdot\lr{\mc{V}\delta z'\delta z'}+\frac{\p}{\p z}\cdot\lr{\mc{V}\delta z\delta z'}}\,dz'dz''dz'''\\
&-\frac{1}{2}\int \mc{F}'\mc{F}_z\cdot\lrs{\frac{\p}{\p z'}\cdot\lr{\mc{V}\delta z'\delta z}+\frac{\p}{\p z}\cdot\lr{\mc{V}\delta z\delta z}}\,dz'dz''dz'''\\
&+\frac{1}{2}\int \frac{\p}{\p z'}\cdot\lrs{\mc{F}\mc{F}'\mc{V}\lr{\delta z'\frac{\p\log \mc{F}'}{\p z'}\cdot\delta z'+\delta z'\frac{\p\log \mc{F}}{\p z}\cdot\delta z}}\,dz'dz''dz'''\\
&+\frac{1}{2}\frac{\p}{\p z}\cdot\int \mc{F}\mc{F}'\mc{V}\lr{\delta z\delta z'\cdot \frac{\p\log \mc{F}'}{\p z'}+\delta z\delta z\cdot \frac{\p\log \mc{F}}{\p z}}\,dz'dz''dz'''\label{Cop2}
}
Under suitable boundary conditions, the fourth term on the right-hand side vanishes. 
In addition, the binary interaction $V$ and the scattering volume density per unit time $\mc{V}$ must be consistent with conservation of particle number, i.e.,
\eq{
\int \mathscr{C}\lr{f,f}\,d\bol{x}dp=0\implies\int\mathscr{C}\lr{f,f}\,dz=0.\label{dndtc0}
}
Dropping boundary integrals, 
we thus obtain the explicit  condition
\eq{
0=&\int \mc{F}\mc{F}_z'\cdot\lrs{\mc{V}\delta z'-\frac{1}{2}\frac{\p}{\p z'}\cdot\lr{\mc{V}\delta z'\delta z'}-\frac{1}{2}\frac{\p}{\p z}\cdot\lr{\mc{V}\delta z\delta z'}}\,dzdz'dz''dz'''\\&+
\int \mc{F}'\mc{F}_z\cdot\lrs{
\mc{V}\delta z-\frac{1}{2}\frac{\p}{\p z'}\cdot\lr{\mc{V}\delta z'\delta z}-\frac{1}{2}\frac{\p}{\p z}\cdot\lr{\mc{V}\delta z\delta z}
}\,dzdz'dz''dz''',
}
which must hold for any $\mc{F}$ and $\mc{F}'$. However,
both $\mc{V}$ and $\delta z$ are properties of binary collisions, implying that the terms within the square brackets are not determined by $\mc{F}$. 
Hence, for the collision process to be consistent with conservation of particle number we should enforce the following requirements on $\mc{V}$:
\eq{
&\int \lrs{\mc{V}\delta z'-\frac{1}{2}\frac{\p}{\p z'}\cdot\lr{\mc{V}\delta z'\delta z'}-\frac{1}{2}\frac{\p}{\p z}\cdot\lr{\mc{V}\delta z\delta z'}}\,dz''dz'''=0,\\&\int \lrs{
\mc{V}\delta z-\frac{1}{2}\frac{\p}{\p z'}\cdot\lr{\mc{V}\delta z'\delta z}-\frac{1}{2}\frac{\p}{\p z}\cdot\lr{\mc{V}\delta z\delta z}
}\,dz''dz'''=0.
}
The collision operator \eqref{Cop2} thus reduces to
\eq{
\mathscr{C}\lr{f,f}=\frac{1}{2}\frac{\p}{\p z}\cdot\int \frac{ff'}{\gamma\gamma'}\mc{V}\lrs{\delta z\delta z'\cdot\frac{\p\log\lr{f'/\gamma'}}{\p z'}+\delta z\delta z\cdot\frac{\p\log \lr{f/\gamma}}{\p z}}\,dz'dz''dz''',\label{Cop3}
}
which can be equivalently written as
\eq{
\mathscr{C}\left(f,f\right)
=\frac{1}{2}
\frac{\p}{\p z}\cdot\int\mc{V}\delta z\lr{\mc{F}\mc{F}'_z\cdot\delta z'+\mc{F}'\mc{F}_z\cdot\delta z}dz'dz''dz'''.
}
Using the phase space displacements \eqref{dz} and \eqref{dzp} derived before, 
we can now define the interaction tensors
\eq{
{\tilde{{\Pi}}_{\mc{V}}}=-\frac{1}{2}\int\frac{\mc{V}}{\gamma\gamma'}\lr{\mathbb{P}\cdot\int_{\tau_c}V_z\,dt}\lr{\mathbb{P}'\cdot\int_{\tau_c}V_z'\,dt}\,dz''dz''',\label{Pip0}
}
and
\eq{
{{\Pi}_{\mc{V}}}=\frac{1}{2}\int\frac{\mc{V}}{\gamma^2}\lr{\mathbb{P}\cdot\int_{\tau_c}V_z\,dt}\lr{\mathbb{P}\cdot\int_{\tau_c}V_z\,dt}\,dz''dz''',\label{Pi0}
}
and express the collision operator \eqref{Cop3} as
\eq{
\mathscr{C}\lr{f,f}=\frac{\p}{\p z}\cdot\lrc{ \frac{f}{\gamma}J_{\Sigma}\cdot\int \frac{f'}{\gamma'}\lrs{\tilde{\Pi}_{\mc{V}}\cdot J_{\Sigma}'\cdot\frac{\p\log\lr{f'/\gamma'}}{\p z'}-\Pi_{\mc{V}}\cdot J_{\Sigma}\cdot\frac{\p\log\lr{f/\gamma}}{\p z}}\,dz'},\label{LECX0}
}
which is the desired 
expression \eqref{LEC}. 
Next, assuming that interactions between colliding particles 
separated by large coordinate-time differences are 
less likely and less consequential, 
we may expand the scattering volume density per unit time 
in powers of the coordinate-time difference 
around the coincident limit $t = t' = t'' = t'''$, 
\eq{
\mc{V}=\mc{V}_t\lr{z,z';z'',z'''}\delta\lr{x^0-x^0{}'}\delta\lr{x^0-x^0{}''}\delta\lr{x^0-x^0{}'''}+O\lr{\epsilon},
}
to obtain the leading order interaction tensors
\eq{
{\tilde{{\Pi}}}=-\frac{1}{2}
\frac{\Gamma_t}{\gamma\gamma'}
\lr{\mathbb{P}\cdot\int_{\tau_c}V_z\,dt}\lr{\mathbb{P}'\cdot\int_{\tau_c}V_z'\,dt}
,\label{Pip}
}
and
\eq{
{{\Pi}}=\frac{1}{2}\frac{\Gamma_t}{\gamma^2}
\lr{\mathbb{P}\cdot\int_{\tau_c}V_z\,dt}\lr{\mathbb{P}\cdot\int_{\tau_c}V_z\,dt}
,\label{Pi}
}
where 
\eq{
\Gamma_t=\int\mc{V}_t\,dV''dV''',
}
is the \ti{scattering frequency}, 
$dV=d\bol{x}dp$,  
and the resulting collision operator \eqref{LECX0} as
\eq{
\mathscr{C}\lr{f,f}=\frac{\p}{\p z}\cdot\lrc{ \frac{f}{\gamma}J_{\Sigma}\cdot\int \frac{f'}{\gamma'}\lrs{\tilde{\Pi}\cdot J_{\Sigma}'\cdot\frac{\p\log\lr{f'/\gamma'}}{\p z'}-\Pi\cdot J_{\Sigma}\cdot\frac{\p\log\lr{f/\gamma}}{\p z}}\,dV'},\label{LECX}
}
where all quantities are now  evaluated at the same coordinate-time $t$. 
We note the symmetry $\Gamma_t\lr{z,z'}=\Gamma_t\lr{z',z}$. 


We conclude this section with a remark on the practical evaluation of the interaction tensors \eqref{Pip} and \eqref{Pi}, and, in particular, on the integral 
\eq{
\int_{\tau_c} V_z\,dt .
}
A direct approximation can be obtained by first solving the ODEs
\eq{
\frac{dz}{dt}=\gamma^{-1}J_{c}\cdot V_z,
\qquad 
\frac{dz'}{dt}=\gamma'{}^{-1}J_{c}\cdot V_{z'},
}
with initial conditions $z(t)$ and $z'(t)$, respectively, to generate the orbits over the interval $[t,t+\tau_c]$. 
The integral can then be evaluated as
\eq{
\int_{t}^{t+\tau_c} V_z\!\left(z(\mf{t}),z'(\mf{t})\right)\,d\mf{t}.
}
In practice, however, this procedure may be inconvenient. 
A simpler approach is to replace the impulse $\int_{\tau_c}V_z\,dt$ with a characteristic value, in analogy with a diffusion coefficient in the context of Fokker–Planck equations. 
As will become evident in Sec.~6, such a simplification does not affect the functional form of the resulting end states (thermodynamic equilibria).

\section{The Landau--Einstein system}
Using the collision operator~\eqref{LEC}, we obtain the Landau--Einstein system~\eqref{LE}: 
\sys{
&\lrc{\frac{f}{\gamma},\mc{H}}=\mathscr{C}\lr{f,f},\label{LEf}\\
& R^{\mu\nu}-\frac{1}{2}Rg^{\mu\nu}=\kappa T^{\mu\nu}.\label{LEg}
}{LE2}
In this section we investigate system~\eqref{LE2} with collision operator \eqref{LECX}, establishing conservation laws and an H-theorem. A detailed discussion of thermodynamic equilibria is given in Sec.~6.

\subsection{Energy-momentum tensor}

We remark that the precise form of the energy–momentum tensor
\begin{equation}
T^{\mu\nu}
= \int \frac{f p^{\mu} p^{\nu}}{m\gamma\sqrt{-\mf g}}\;dp
\;+\; F^{\mu\nu} \;+\; C^{\mu\nu},
\label{Tmn}
\end{equation}
depends on the chosen binary interaction potential \(V\).
The first (kinetic) term on the right-hand side of \eqref{Tmn}
corresponds to the free-streaming (interaction-free) geodesic limit,
i.e., the collisionless Vlasov–Einstein system, whereas \(F^{\mu\nu}\)
and \(C^{\mu\nu}\) collect the field and collisional contributions, respectively. 
One can verify that the (interaction-free) Vlasov-Einstein system
\sys{
&\lrc{\frac{f}{\gamma},H}= \frac{\p f}{\p t}+\frac{\p}{\p x^j}\lr{\frac{p^j}{m\gamma}f}-\frac{\p}{\p p_{\mu}}\lr{\frac{1}{2m\gamma}\frac{\p g^{\lambda\nu}}{\p x^{\mu}}p_{\lambda}p_{\nu}f}=0,\\
&R^{\mu\nu}-\frac{1}{2}Rg^{\mu\nu}=\frac{8\pi G}{mc^4}\int \frac{fp^{\mu}p^{\nu}}{\gamma\sqrt{-\mf{g}}}dp.
}{EVgeo}
is consistent with the requirement that the Einstein tensor $G^{\mu\nu}=R^{\mu\nu}-\frac{1}{2}Rg^{\mu\nu}$ is divergence--free. 
Recall that the action of the covariant derivative $\nabla_{\mu}$ on vector fields $Y=Y^{\nu}\p_{\nu}$ and contravariant $2$-tensors $T^{\nu\lambda}$ can be expressed as follows:
\sys{
&\nabla_{\mu}Y^{\nu}=\frac{\p Y^{\nu}}{\p x^{\mu}}+\Gamma_{\mu\alpha}^{\nu}Y^{\alpha},\\
&\nabla_{\mu}T^{\nu\lambda}=\frac{\p T^{\nu\lambda}}{\p x^{\mu}}+\Gamma^{\lambda}_{\mu\alpha} T^{\nu\alpha}+\Gamma^{\nu}_{\mu\alpha} T^{\alpha\lambda},
}{cov}
With $\mc{F}=f/\gamma$,  it follows that
\eq{
\nabla_{\mu}\lr{\frac{\mc{F}p^{\mu}p^{\nu}}{\sqrt{-\mf{g}}}}=&\nabla_{\mu}\lr{\frac{\mc{F}}{\sqrt{-\mf{g}}}p^{\mu}}p^{\nu}+\frac{\mc{F}}{\sqrt{-\mf{g}}}p^{\mu}\nabla_{\mu}p^{\nu}\\
=&\frac{\p}{\p x^{\mu}}\lr{\frac{\mc{F}}{\sqrt{-\mf{g}}}p^{\mu}}p^{\nu}+\frac{\mc{F}}{\sqrt{-\mf{g}}}\Gamma_{\mu\alpha}^{\mu}p^{\alpha}p^{\nu}+\frac{\mc{F}}{\sqrt{-\mf{g}}}p^{\mu}\frac{\p g^{\nu\alpha}}{\p x^{\mu}}p_{\alpha}+\frac{\mc{F}}{\sqrt{-\mf{g}}}p^{\mu}\Gamma_{\mu\alpha}^{\nu}p^{\alpha}\\
=&\frac{1}{\sqrt{-\mf{g}}}\frac{\p}{\p x^{\mu}}\lr{\mc{F}p^{\mu}}p^{\nu}+\frac{\mc{F}}{\mf{g}}\frac{\p\sqrt{-\mf{g}}}{\p x^{\mu}}p^{\mu}p^{\nu}+\frac{\mc{F}}{\sqrt{-\mf{g}}}\frac{\p\log\sqrt{-\mf{g}}}{\p x^{\alpha}}p^{\alpha}p^{\nu}-\frac{\mc{F}}{\sqrt{-\mf{g}}}g^{\nu\beta}\frac{\p g_{\alpha\beta}}{\p x^{\mu}}p^{\alpha}p^{\mu}\\&+\frac{\mc{F}}{2\sqrt{-\mf{g}}}g^{\nu\beta}\lr{\frac{\p g_{\mu\beta}}{\p x^{\alpha}}+\frac{\p g_{\alpha\beta}}{\p x^{\mu}}-\frac{\p g_{\mu\alpha}}{\p x^{\beta}}}p^{\mu}p^{\alpha}\\
=&\frac{1}{\sqrt{-\mf{g}}}\frac{\p}{\p x^{\mu}}\lr{\mc{F}p^{\mu}}p^{\nu}-\frac{\mc{F}}{2\sqrt{-\mf{g}}}g^{\nu\beta}\frac{\p g_{\mu\alpha}}{\p x^{\beta}}p^{\mu}p^{\alpha}.\label{EVgeo0}
}
On the other hand, notice that 
the Vlasov-Einstein system \eqref{EVgeo} can be equivalently expressed as
\sys{
&\frac{\p}{\p x^{\mu}}\lr{\frac{p^{\mu}}{m}\mc{F}}=-\frac{\p}{\p p_{\mu}}\lr{\dot{p}_{\mu}\mc{F}},\\
&R^{\mu\nu}-\frac{1}{2}Rg^{\mu\nu}=\frac{8\pi G}{mc^4}\int \frac{\mc{F}p^{\mu}p^{\nu}}{\sqrt{-\mf{g}}}\,dp.
}{RelVlas3}
Integrating equation \eqref{EVgeo0} in momentum space, and neglecting boundary terms, we therefore obtain
\eq{
\int\nabla_{\mu}\lr{\frac{\mc{F}p^{\mu}p^{\nu}}{\sqrt{-\mf{g}}}}\,dp=&\int\frac{\mc{F}}{\sqrt{-\mf{g}}}\lr{m\dot{p}_{\mu}\frac{\p p^{\nu}}{\p p_{\mu}}-\frac{1}{2}g^{\nu\beta}\frac{\p g_{\mu\alpha}}{\p x^{\beta}}p^{\mu}p^{\alpha}}dp\\
=&-\int\frac{\mc{F}}{2\sqrt{-\mf{g}}}\lr{g^{\nu\mu}\frac{\p g^{\alpha\beta}}{\p x^{\mu}}p_{\alpha }p_{\beta}+g^{\nu\beta}\frac{\p g_{\mu\alpha}}{\p x^{\beta}}p^{\mu}p^{\alpha}}dp\\
=&\int\frac{\mc{F}}{2\sqrt{-\mf{g}}}\lr{g^{\nu\mu}\frac{\p g_{\alpha\beta}}{\p x^{\mu}}-{g}^{\nu\mu}\frac{\p g_{\alpha\beta}}{\p x^{\mu}}}p^{\alpha}p^{\beta}\,dp=0.
}
By the (contracted) Bianchi identity, $\nabla_\mu G^{\mu\nu}=0$, the interaction potential $V$ and the tensors $F^{\mu\nu}$ and $C^{\mu\nu}$ must be chosen so that the total stress--energy is symmetric and  divergence--free, $T^{\mu\nu}=T^{\nu\mu}$ and $\nabla_\mu T^{\mu\nu}=0$. For example, for electromagnetic interactions one may take $F^{\mu\nu}$ to be the Maxwell energy-momentum tensor. Nevertheless, the explicit forms of $F^{\mu\nu}$ and $C^{\mu\nu}$ will not be needed to establish the conservation laws, the H-theorem, or the characterization of thermodynamic equilibria for the Landau–Einstein system.

\subsection{Conservation laws}
Here, we examine the conservation laws of the Landau–Einstein system \eqref{LE} with respect to the coordinate-time $t$. Specifically, we have
\begin{proposition}
The Landau–Einstein system \eqref{LE} with collision operator \eqref{LECX}  preserves the total particle number 
\eq{
N=\int f\,d\bol{x}dp, 
}
and the total energy
\eq{
\mathscr{H}
=&\int f H\,d\bol{x}dp
+\int ff' V\,d\bol{x}dp dz'+\int ff'V\,dzd\bol{x}'dp'+\int f'H'\,d\bol{x}'dp'\\
=&\int f\lr{H+\Phi}\,d\bol{x}dp
+\int f'\lr{H'+\Phi'}\,d\bol{x}'dp'\\
=&\int f\mc{H}\,d\bol{x}dp+\int f'\mc{H}'\,d\bol{x}dp=2U_t,
\label{HT}
}
with respect to coordinate-time $t$.
\end{proposition}
Notice that the two integrals in the last expression of \eqref{HT} are identical, each equal to the functional $U_t$ defined in eq. \eqref{Ut}; we keep this symmetrized form to simplify the calculation of $d\mathscr{H}/dt$.
We also remark that, for notational simplicity, the points at which the integrands are evaluated are left implicit, and that $\mathscr{H}$ is a function of the time parameter $t$ alone.

\begin{proof}
First, let us consider the total particle number $N$. 
We have
\eq{
\frac{dN}{dt}=\int \lrs{-\frac{\p}{\p x^i}\lr{\frac{f}{\gamma}\frac{\p\mc{H}}{\p p_i}}+\frac{\p}{\p p_{\mu}}\lr{\frac{f}{\gamma}\frac{\p\mc{H}}{\p x^{\mu}}}+\mathscr{C}\lr{f,f}}\,d\bol{x}dp.
}
Neglecting boundary contributions, 
denoting with 
\eq{
Z_c=J_{\Sigma}\cdot\int \frac{f'}{\gamma'}\lrs{\tilde{\Pi}\cdot J_{\Sigma}'\cdot\frac{\p\log\lr{f'/\gamma'}}{\p z'}-\Pi\cdot J_{\Sigma}\cdot\frac{\p\log\lr{f/\gamma}}{\p z}}\,d\bol{x}'dp'\label{Zc}
}
the total collisional phase space velocity, 
and recalling that the phase-space displacements satisfy $\delta x^0=0\implies Z^0_c=0$, which removes the term $\frac{\p}{\p x^0}\!\lr{\frac{f}{\gamma}Z^0_c}$ from the collision operator \eqref{LECX}, we can write
\eq{
\mathscr{C}\!\lr{f,f}
=\frac{\p}{\p z^A}\!\lr{\frac{f}{\gamma}Z^A_c}
=\frac{\p}{\p x^i}\!\lr{\frac{f}{\gamma}Z^i_c}
+\frac{\p}{\p p_{\mu}}\!\lr{\frac{f}{\gamma}Z^{4+\mu}_c},
}
and thus obtain
\eq{
\frac{dN}{dt}
=\int \mathscr{C}\lr{f,f}\,d\bol{x}dp
=0.
}
We now move to $\mathscr{H}$, and recall the notation  $dV=d\bol{x}dp$ for the measure on the reduced phase space $\lr{\bol{x},p}$. 
The rate of change of $\mathscr{H}$ is 
\eq{
\frac{d\mathscr{H}}{dt}=\int
\frac{\p f}{\p t}\mc{H}\,dV+\int f \frac{\p \mc{H}}{\p t}\,dV+\int \frac{\p f'}{\p t}\mc{H}'\,dV'
+\int f'\frac{\p \mc{H}'}{\p t}\,dV'.
}
Neglecting boundary terms, it follows that
\eq{
\frac{d\mathscr{H}}{dt}
=&\int \lrs{
-\frac{\p}{\p x^i}
\lr{\frac{f}{\gamma}\frac{\p\mc{H}}{\p p_i}}
+\frac{\p}{\p p_{\mu}}
\lr{
\frac{f}{\gamma}
\frac{\p\mc{H}}{\p x^{\mu}}
}
+\mathscr{C}\lr{f,f}
}
\mc{H}\,dV+\int f\frac{\p\mc{H}}{\p t}\,dV\\
&+\int \lrs{
-\frac{\p}{\p x^i{}'}
\lr{\frac{f'}{\gamma'}\frac{\p\mc{H}'}{\p p_i'}}
+\frac{\p}{\p p_{\mu}'}
\lr{
\frac{f'}{\gamma'}
\frac{\p\mc{H}'}{\p x^{\mu}{}'}
}
+\mathscr{C}\lr{f,f}'
}
\mc{H}'\,dV'+\int f'\frac{\p\mc{H}'}{\p t}\,dV'\\
=&-\int \frac{f}{\gamma}\frac{\p\mc{H}}{\p x^0}\frac{\p\mc{H}}{\p p_0}\,dV+\int f\frac{\p\mc{H}}{\p t}\,dV+\int\mathscr{C}\lr{f,f}\mc{H}\,dV\\
&-\int \frac{f'}{\gamma'}\frac{\p\mc{H}'}{\p x^0{}'}\frac{\p\mc{H}'}{\p p_0}\,dV'+\int f\frac{\p\mc{H}'}{\p t}\,dV'+\int\mathscr{C}\lr{f,f}'\mc{H}'\,dV'.
}
Recalling that $c\gamma =\p\mc{H}/\p p_0$, that $Z^0=0$,  using the expression of the collision operator \eqref{LECX}, and integrating by parts, we arrive at
\eq{
\frac{d\mathscr{H}}{dt}=&\int\mathscr{C}\lr{f,f}\mc{H}\,dV+\int\mathscr{C}\lr{f,f}'\mc{H}'\,dV'\\
=&-\int \frac{f}{\gamma}Z_c\cdot\frac{\p\mc{H}}{\p z}\,dV-\int \frac{f'}{\gamma'}Z'_c\cdot\frac{\p\mc{H}'}{\p z'}\,dV'\\
=&
\int \frac{ff'}{\gamma\gamma'}\lr{\tilde{\Pi}\cdot J_{\Sigma}'\cdot\frac{\p\log\mc{F}'}{\p z'}-\Pi\cdot J_{\Sigma}\cdot\frac{\p\log \mc{F}}{\p z}}\cdot J_{\Sigma}\cdot\frac{\p\mc{H}}{\p z}\,dVdV'
\\
&+\int \frac{ff'}{\gamma\gamma'}\lr{
\tilde{\Pi}'\cdot J_{\Sigma}\cdot\frac{\p\log\mc{F}}{\p z}-\Pi'\cdot J_{\Sigma}'\cdot\frac{\p\log \mc{F}'}{\p z'}}\cdot J_{\Sigma}'\cdot\frac{\p\mc{H}'}{\p z'}\,dVdV'.\label{dHdt0}
}
Now note that the interaction tensors \eqref{Pip} and \eqref{Pi} have the following forms:
\eq{
\tilde{\Pi}=-AA',\qquad \Pi=AA,\qquad \tilde{\Pi}'=-A'A,\qquad \Pi'=A'A',\label{AA}
}
with 
\eq{
A=\frac{1}{\gamma}\sqrt{\frac{\Gamma_t}{2}}\mathbb{P}\cdot a,\qquad a=\int_{\tau_c}V_z\,dt.
}
Equation \eqref{dHdt0} thus becomes
\eq{
\frac{d\mathscr{H}}{dt}=&-\int \frac{ff'}{\gamma\gamma'}\lr{A'\cdot J_{\Sigma}'\cdot\frac{\p\log\mc{F}'}{\p z'}
+A\cdot J_{\Sigma}\cdot\frac{\p\log\mc{F}}{\p z}
}\lr{\gamma A\cdot\xi+\gamma' A'\cdot\xi'}\, dVdV',
}
where we used equation \eqref{xi}. 
Recalling the definitions \eqref{q}, \eqref{delta}, and \eqref{mathbbP}, we  obtain 
\eq{
 \lr{\gamma A\cdot\xi+\gamma' A'\cdot\xi'}\sqrt{\frac{2}{\Gamma_t}}=&
\lr{\mathbb{P}^{\perp}_q\cdot a-\frac{\Delta\cdot\xi'}{\abs{q}^2}q^{\flat}}\cdot\lr{q+\xi'}+\lr{\mathbb{P}^{\perp}_q\cdot a'-\frac{\Delta\cdot\xi}{\abs{q}^2}q^{\flat}{}'}\cdot\xi'\\
=&\lrs{\mathbb{P}^{\perp}_q\cdot a-\Delta-\frac{\Delta\cdot\xi'}{\abs{q}^2}q^{\flat}+\mathbb{P}^{\perp}_q\cdot\lr{\Delta-a}-\frac{\Delta\cdot\xi}{\abs{q}^2}q^{\flat}{}'}\cdot\xi'\\
=&\lr{\Delta-\Delta}\cdot\xi'=0.\label{Axi}
}
We have thus shown that the total energy $\mathscr{H}$ is constant, 
\eq{
\frac{d\mathscr{H}}{dt}=0.
}
\end{proof}

\begin{remark}
The total energy of the classical Vlasov equation \cite{PJMVM}
\eq{
\mathscr{H}_V=\int f\lr{t,\bol{x},\bol{p}}\lr{\frac{\bol{p}^2}{2m}+\frac{1}{2}\,\Phi_{C}\lr{t,\bol{x}}}\,d\bol{x}\,d\bol{p},
}
where $\Phi_C$ is the electrostatic potential, carries a factor $1/2$ in front of the potential term. This factor prevents double counting of pairwise Coulomb interactions (already evident in the discrete two-particle energy $\bol{p}_1^{\,2}/2m+V_C+\bol{p}_2^{\,2}/2m$). By contrast, in the $8$-dimensional Landau–Einstein system $\mathscr{H}$ is conserved because the $p_0$-flux cancels the explicit $\p\mathcal{H}/\p t$; after projection to $6$ dimensions $\lr{\bol{x},\bol{p}}$ that cancellation is gone, and the conserved energy becomes the symmetric quadratic form $\mathscr{H}_V$. 
\end{remark}

\subsection{H-theorem}
As explained in Sec.~2.3, although a coordinate-time independent invariant measure on the $7$-dimensional reduced phase space $\lr{\bol{x},p}$ is generally unavailable, the weighted measure $\gamma\,dV=\gamma\,d\bol{x}dp$ is compatible with the conservation of the entropy functional $S_t$ (see eq.~\eqref{St} with $J=\gamma$) in the collisionless case. 
Here, we show that in the presence of collisions the (symmetrized) entropy functional
\eq{
\mathscr{S}=-\int f\log\!\lr{\frac{f}{\gamma}}\,dV-\int f'\log\!\lr{\frac{f'}{\gamma'}}\,dV' \;=\; 2S_t,
}
has a nonnegative time derivative under the usual assumption of vanishing/decaying boundary contributions. 
In other words, the model recovers the monotonicity expected from the second law of thermodynamics.

\begin{proposition}
The Landau--Einstein system \eqref{LE}, with collision operator \eqref{LECX}, satisfies the $H$-theorem
\eq{
\frac{d\mathscr{S}}{dt}\ge 0,
}
under vanishing boundary conditions.
\end{proposition}

\begin{proof}
We begin by evaluating the rate of change of $S_t$ due to the Landau--Einstein system \eqref{LE} with collision operator \eqref{LECX}: 
\eq{
\frac{dS_t}{dt}=&-\int \frac{\p f}{\p t}\lr{1+\log\mc{F}}\,dV+\int f\frac{\p\log\gamma}{\p t}\,dV\\
=&\int\lrs{
\frac{\p}{\p x^i}\lr{\mc{F}\frac{\p\mc{H}}{\p p_i}}-\frac{\p}{\p p_{\mu}}\lr{\mc{F}\frac{\p\mc{H}}{\p x^{\mu}}}
-\mathscr{C}\lr{f,f}}\lr{1+\log\mc{F}}\,dV
+\int \mc{F}\frac{\p^2\mc{H}}{\p x^0\p p_0}\,dV\\
=&\int\lr{\frac{\p\mc{F}}{\p p_0}\frac{\p\mc{H}}{\p x^0}+\mc{F}\frac{\p^2\mc{H}}{\p x^0\p p_0}}\,dV+\int \mc{F}Z_c\cdot\frac{\p\log\mc{F}}{\p z}\,dV\\
=&\int \mc{F}Z_c\cdot\frac{\p\log\mc{F}}{\p z}\,dV.\label{dStdtc}
}
Recalling the expression of the collisional phase space velocity $Z_c$ (see eq. \eqref{Zc}) and the decomposition of the interaction tensors therein (see eq. \eqref{AA}), it follows that
\eq{
\frac{d\mathscr{S}}{dt}=&\int \mc{F}Z_c\cdot\frac{\p\log\mc{F}}{\p z}\,dV+\int \mc{F}'Z_c'\cdot\frac{\p\log\mc{F}'}{\p z'}\,dV'\\
=&\int \mc{F}\mc{F}'\lr{A'\cdot J_{\Sigma}'\cdot\frac{\p\log \mc{F}'}{\p z'}+A\cdot J_{\Sigma}\cdot\frac{\p\log\mc{F}}{\p z}}A\cdot J_{\Sigma}\cdot\frac{\p\log\mc{F}}{\p z}\,dVdV'\\
&+\int\mc{F}\mc{F}'\lr{A\cdot J_{\Sigma}\cdot\frac{\p\log\mc{F}}{\p z}+A'\cdot J_{\Sigma}'\cdot\frac{\p\log \mc{F}'}{\p z'}} A'\cdot J_{\Sigma}'\cdot\frac{\p\log\mc{F}'}{\p z'}\,dVdV'\\
=&
\int\mc{F}\mc{F}'\lr{A\cdot J_{\Sigma}\cdot\frac{\p\log\mc{F}}{\p z}+A'\cdot J_{\Sigma}'\cdot\frac{\p\log \mc{F}'}{\p z'}}^2\,dVdV'.\label{Hthm}
}
Assuming that $\mc{F},\mc{F}'\geq 0$ at all times $t$, we have thus shown that
\eq{
\frac{d\mathscr{S}}{dt}\geq 0.
}
\end{proof}


\section{Thermodynamic equilibrium}
In this section, we analyze the equilibria of the Landau--Einstein system \eqref{LE} with collision operator \eqref{LECX} and discuss their Minkowski and classical limits.

Let $f_{\infty}\lr{x,p}$ denote the equilibrium distribution function, and set $\mc{F}_{\infty}=f_{\infty}/\gamma$. 
Note that $f_{\infty}$ may, in general, depend on $x^0$, as determined by the adopted notion of equilibrium (cf.~Table~1). 
For any physically acceptable end state we require
\eq{
f_{\infty}=\gamma\,\delta\lr{\chi}\,w\lr{x,p},\label{fL0}
}
where $\delta\lr{\chi}$ enforces the constraint $\chi=0$, ensuring that all particles lie on the constant speed of light submanifold, and $w$ is some phase space function to be determined from the equilibrium conditions.

\subsection{Liouville equilibria}
The first nontrivial class of equilibria 
are Liouville type (remember table 1), where both 
ideal and collision terms of the kinetic equation for the distribution function $f$ vanish independently. 
As clear from equation \eqref{dStdtc}, a necessary condition for the collision term $\mathscr{C}{\lr{f,f}}$ 
to vanish is that $dS_t/dt=d\mathscr{S}/dt=0$. 
Hence, the quadratic term within the integrand of eq. \eqref{Hthm} must be identically zero,
\eq{
&A\cdot J_{\Sigma}\cdot\frac{\p\log\mc{F}_{\infty}}{\p z}+A'\cdot J_{\Sigma}'\cdot\frac{\p\log \mc{F}'_{\infty}}{\p z'}=0.
}
If $\log\mc{F}_{\infty}=-\beta\mc{H}$ for some $\beta\in\mathbb{R}$, we have $A\cdot J_{\Sigma}\cdot\p\log\mc{F}/\p z=-\beta\gamma A\cdot\xi$. 
We have already seen (cf. eq. \eqref{Axi}) that 
$\gamma A\cdot\xi+\gamma'A'\cdot\xi'$ identically vanishes. 
Furthermore, the tensor $J_{\Sigma}$ has a nontrivial null-space spanned by $dC\lr{\chi,t}$, where $C\lr{\chi,t}$ is any function of $\chi$ and $t$. 
A candidate Liouville equilibrium is therefore given by
\eq{
f_{\infty}=\gamma\exp\lrc{-\beta\mc{H}+C},
}
which annihilates the collision term $\mathscr{C}\lr{f_{\infty},f_{\infty}}=0$. 
On the other hand, for such $f_{\infty}$ to define
a Liouville equilibrium the ideal term, 
\eq{
\lrc{\frac{f_{\infty}}{\gamma},\mc{H}}=\exp\lrc{-\beta\mc{H}+C}\lr{\frac{\p C}{\p\chi}\lrc{\chi,\mc{H}}+\frac{\p C}{\p t}\lrc{t,\mc{H}}},\label{id0}
}
must also vanish. 
Now observe that $\lrc{t,\mc{H}}=\p\mc{H}/\p p_0\neq 0$ in general. Hence, we set $\p C/\p t=0$, or $C=C\lr{\chi}$. 
In addition, for any speed of light compatible  Hamiltonian $\mc{H}$, eq. \eqref{slc}, we have $\lrc{\chi,\mc{H}}=0$, leading to the desired result
\eq{
\lrc{\frac{f_{\infty}}{\gamma},\mc{H}}=0.
}
On the other hand, particles lie on the submanifold $\chi=0$ provided that  
$e^C=\delta\lr{\chi}\ell\lr{\chi}$ 
for some function $\ell\lr{\chi}$. 
However, integration of $\delta$ results in a constant $\ell\lr{0}$ that can be absorbed in the normalization of the distribution. Hence, we set $\ell=1/Z\in\mathbb{R}_{\geq 0}$, and arrive at the Liouville equilibrium distribution function 
\eq{
f_{\infty}=\frac{1}{Z}\gamma\,\delta\lr{\chi}\,\exp\lrc{-\beta\mc{H}}.\label{Leq} }
It is useful to evaluate the spatial density associated with \eqref{Leq},
\eq{
\rho\lr{x}=\int f_{\infty}\,dp=
\frac{1}{Z}\int\lrs{\gamma\exp\lrc{-\beta\mc{H}}\frac{\p p_0}{\p\chi}}_{\chi=0}\,d\bol{p}.
}
Interestingly, when 
the average interaction energy vanishes $\Phi=0$, 
$\mc{H}=m\chi/2$ is identified with the mass--shell constraint, and the metric is spacetime orthogonal $g_{0i}=0$, $i=1,2,3$, we obtain 
\eq{
\frac{\partial p_0}{\partial \chi} = \frac{m}{2c\gamma},
}
so that
\eq{
\rho = \frac{m}{2cZ}\int d\bol{p}.
\label{Leqr}
}
The 
constant 
density \eqref{Leqr} reflects the fact that, in this setting, the only constraint affecting particle dynamics is the speed of light limit, with no nontrivial collisional invariants. As a consequence, the system undergoes complete thermalization once the Liouville state is reached. 
We emphasize, however, that in general $\mc{H}\neq \mc{H}\lr{\chi}$, and the Liouville equilibrium \eqref{Leq} encodes the nontrivial (grazing) invariant $\mc{H}$.
 Statistical distributions of the form \eqref{Leq} have been proposed on the basis of maximum entropy arguments~\cite{SatoCGQ1}.

A remarkable feature of the present construction is that 
in the classical limit 
the Liouville equilibrium distribution function \eqref{Leq} reduces to a Maxwell-Boltzmann distribution. 

\begin{proposition}
In the classical limit, the Liouville equilibrium distribution function \eqref{Leq} reduces to a Maxwell-Boltzmann distribution. 
\end{proposition}

\begin{proof}
First, we must determine how the 
speed of light constraint $\chi=0$ is modified by the classical limit $c\rightarrow+\infty$. 
Noting that $g\rightarrow\eta$ in this limit, we have 
\eq{
\lim_{c\rightarrow+\infty}\chi=\lim_{c\to  +\infty}\chi_c,\qquad \chi_c=c^2\lr{1-\gamma^2}=c^2-\lr{\frac{\p\mc{H}}{\p p_0}}^2=c^2-\lr{\frac{p^0}{m}+\frac{\p\Phi}{\p p_0}}^2.
}
Assuming $\gamma>0$, it follows that
\eq{
\chi_c=0\iff \frac{p^0}{m}=c-\frac{\p\Phi}{\p p_0}\iff \gamma=1.
}
Then, the $p_0$-averaged distribution function can be evaluated as follows: 
\eq{
\int f_{\infty}dp_0=&\frac{1}{Z}\int\delta\lr{\chi_c}\exp\lrc{-\beta\lr{H+\Phi}}\,dp_0\\=&
\frac{1}{Z}\int\frac{\delta\lr{\chi_c}\exp\lrc{-\beta\lr{H+\Phi}}}{2\lr{\frac{p^0}{m}+\frac{\p\Phi}{\p p_0}}\lr{\frac{1}{m}-\frac{\p^2\Phi}{\p p_0^2}}}\,d\chi_c\\
=&\frac{m}{2cZ}\lr{1-m\frac{\p^2\Phi}{\p p_0^2}}_{\chi_c=0}^{-1}\exp\lrc{-\frac{1}{2}\beta mc^2-\beta\frac{\bol{p}^2}{2m}-\beta\lrs{\Phi-\frac{1}{2} m\lr{c-\frac{\p\Phi}{\p p_0}}^2}_{\chi_c=0}}\\
=&
\frac{m}{2cZ}\lr{1-m\frac{\p^2\Phi}{\p p_0^2}}_{\chi_c=0}^{-1}\exp\lrc{-\beta\frac{\bol{p}^2}{2m}-\beta\lrs{
m c\frac{\p\Phi}{\p p_0}-\frac{1}{2}m\lr{\frac{\p\Phi}{\p p_0}}^2+\Phi}_{\chi_c=0}}.\label{MB0}
}
Note that this distribution function contains the Maxwellian weight 
$\exp\lrc{-\beta\bol{p}^2/2m}$, where $\bol{p}^2=p_i p^i$, 
while the additional $\Phi$-dependent factors reflect the specific properties of the interaction. 
\end{proof}

Proposition 6.1 clearly shows that, in the classical limit, one does not need to rely on a coordinate-time Killing symmetry or on the additive conservation of $p_0$ to obtain a Maxwell–Boltzmann distribution at equilibrium. 
It is therefore useful to consider specific examples of \eqref{MB0}. 

For the standard electromagnetic case \eqref{EM}, one has 
\eq{
\frac{\partial \Phi}{\partial p_0}=-\frac{qA^0}{m}, 
\qquad 
\frac{\partial^2 \Phi}{\partial p_0^2}=0.
}
Equation \eqref{MB0} thus becomes
\eq{
\int f_{\infty}\,dp_0
= \frac{m}{2cZ}\exp\lrc{-\beta\frac{\lr{\bol{p}-q\bol{A}}^2}{2m}}.
}
It should not be surprising that the exponential factor of this distribution does not contain the electrostatic term $-\beta qcA^0$, since the underlying additive grazing invariant is $\mc{H}$ rather than $p_0$. 
Indeed, the standard electromagnetic Hamiltonian with $\Phi$ given by \eqref{EM} does not admit ``mean fields'' in the sense that the electromagnetic potentials merely shift the momenta, $p_{\mu}\to p_{\mu}-qA_{\mu}$.

On the other hand, if we consider Vlasov-type average interaction potentials $\Phi$ such that $\partial\Phi/\partial p_0=0$, then the $p_0$-averaged equilibrium distribution in the classical limit takes the familiar form
\eq{
\int f_{\infty}\,dp_0
= \frac{m}{2cZ}\exp\lrc{-\beta\lr{\frac{\bol{p}^2}{2m}+\Phi}}.
}

\subsection{Thermodynamic equilibria}
A thermodynamic equilibrium may be regarded as a Liouville equilibrium in which a Killing coordinate-time symmetry has emerged for the metric and the other physical observables.
Evidently, the Liouville equilibrium \eqref{Leq} obtained in section 6.1 qualifies as a thermodynamic equilibrium of the Landau--Einstein system \eqref{LE} with collision operator \eqref{LECX}, provided that  
$\p f_{\infty}/\p t=0$ and $\p g/\p t=0$ are additionally satisfied. 

Next, let us consider a situation in which
\eq{
\frac{\p g}{\p t}=0,\qquad
\frac{\p \Phi}{\p t}=0.
}
holds.  
Then, $\lrc{p_0,\mc{H}}=-c^{-1}\p\Phi/\p t=0$.   Furthermore, let us assume that collisions are $p_0$-grazing (i.e. $p_0$  is approximately conserved after a binary scattering because, for example, $\p V/\p t=O\lr{\epsilon}$ is small due to the specific form of the interaction potential energy $V$ and the coordinate-time Killing symmetry of the metric $g$; recall the discussion in the last paragraph of section 4.1). 
Then, we may replace the projector \eqref{Proj} appearing in the expression of the tensor  $J_{\Sigma}=\mathbb{P}_{\Sigma}\cdot J_c\cdot\mathbb{P}_{\Sigma}$ with the projector $\mathbb{P}_{\Sigma'}$ on the
$5$-dimensional distribution
\eq{
\Sigma'_z=\bigl\{\delta z\in T_z\!\left(T^{\ast}M\right):d\chi(\delta z)=0,\; dx^0(\delta z)=0,\;dp_0\lr{\delta z}=0\bigr\}.
}
If we further modify the collision operator \eqref{LECX} by replacing $J_{\Sigma}$ with $J_{\Sigma'}=\mathbb{P}_{\Sigma'}\cdot J_c\cdot \mathbb{P}_{\Sigma'}$, one finds 
that 
the Landau--Einstein distribution \eqref{LEd},  
\eq{
f_{\infty}=\frac{1}{Z}\gamma\,\delta\lr{\chi}\exp\lrc{-\beta\mc{H}}\zeta\lr{p_0},\label{LEd2}
}
is a thermodynamic equilibrium of the Landau--Einstein system (with collision operator modified via $J_{\Sigma'}$). 

\begin{remark}
In the relativistic kinetic theory of a simple gas \cite{Israel}, 
molecules physically collide via elastic scattering in a locally Minkowskian frame, 
where the components of the four-momentum 
$p_{\mu}=m g_{\mu\nu}\,dx^{\nu}/d\tau$ 
are additively preserved, and no interaction potential energy $V$ is at play 
(or, equivalently, the elastic scattering effective potential energy has the form 
$V\lr{x-x',p,p'}$ and simultaneously satisfies a vanishing ensemble average, $\Phi=0$). 
Furthermore, because in such a setting $\mc{H}=m\chi/2=H$ is the geodesic Hamiltonian, 
the collision operator \eqref{LECX} can be applied by replacing $\mc{H}$ with $p_0$ 
as the grazing invariant, and by using the purely geodesic Hamiltonian $H$ 
as generator of the ideal part of the dynamics, i.e.  
$\lrc{f/\gamma,H}=\mathscr{C}\lr{f,f}$. 
When $g_{\mu\nu}=\eta_{\mu\nu}$ is the Minkowski metric, the 
corresponding thermodynamic equilibrium becomes
\eq{
f_{\infty}=\frac{1}{Z}\gamma\,\delta\lr{\chi}\exp\lrc{\beta^0 p_0}
=\frac{1}{Z}\gamma\,\delta\lr{\chi}\exp\lrc{-\beta^0 mc\gamma}, 
\qquad \beta^0\in\mathbb{R}_{\geq 0}, \label{SJd}
}
which is the Synge--J\"uttner distribution. 
Of course, if all the collisional additive invariants $p_{\mu}$ are taken into account, 
then $\beta^0 p_{0}$ should be replaced with $\beta^{\mu}p_{\mu}$ in \eqref{SJd}. 
\end{remark}


\begin{remark}
Note that, however, the Synge-J\"uttner distribution \eqref{SJd} does not include the mean field $\Phi$,  i.e. the non-kinetic Boltzmann weight that would appear with, for example,  Coulomb collisions. 
This contribution automatically appears in $p_0=mg_{0\mu}\lr{\dot{x}^{\mu}-\p\Phi/\p p_{\mu}}$ once the assumption $\Phi=0$ is relaxed, and only $\p \Phi/\p t=0$ is required, ensuring that $\lrc{p_0,\mc{H}}=0$. Then, the non-kinetic Boltzmann weight is given by $g_{0\mu}\p\Phi/\p p_{\mu}$. 
In the familiar electromagnetic case \eqref{EM} 
one obtains $g_{0\mu}\p\Phi/\p p_{\mu}=-qA_{0}/m$. The resulting general relativistic Maxwell-Boltzmann distribution then reads
\eq{
f_{\infty}=
\frac{1}{Z}\gamma\,\delta\lr{\chi}\exp\lrc{-\beta^0\lrs{
mcg_{00}\gamma +qA_0
+mg_{0i}
\dot{x}^i
}.
}
}
\end{remark}

\begin{remark}
In a curved spacetime geometry, 
a covariant interaction potential energy generally does not have the form $V\lr{x-x',p,p'}$. 
Indeed, the covariant notion of distance is encoded in the metric-dependent Synge's world function $\sigma\lr{x,x'}$. 
Hence, one cannot generally expect $p_{\mu}$ to be a collisional additive invariant.  
\end{remark}

\begin{remark} 
As in the case of Liouville equilibria (see proposition 6.1), the distribution function \eqref{LEd2} reduces to a Maxwell-Boltzmann type distribution in the classical limit.
\end{remark}

We conclude this section with Fig.~2, which illustrates the structure of 
general relativistic equilibria emerging from the present study.

\begin{figure}
\centerline{\includegraphics[scale=0.2]{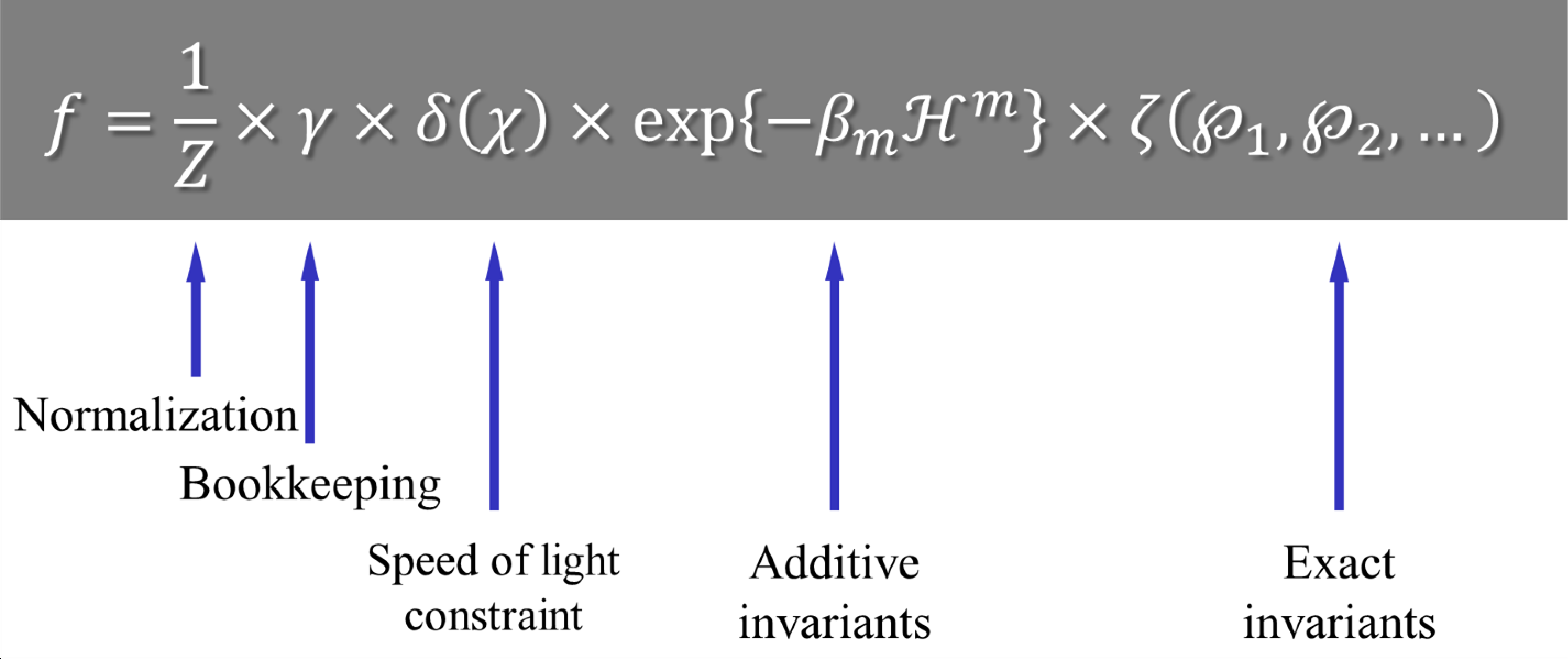}}
\caption{
Structure of general relativistic equilibria: 
the equilibrium distribution comprises a normalization factor $Z^{-1}$, 
the Lorentz factor (reflecting the bookkeeping time parameter $t$), 
a $\delta$-function enforcing the speed of light constraint, 
and an exponential weight determined by the additive invariants $\mc{H}^m$ 
with constant multipliers $\beta_m$, which are additively preserved by collisions 
and exactly preserved by the Poisson bracket, $\lrc{\mc{H}^m,\mc{H}}=0$. 
In addition, a function $\zeta$ depends on the exact invariants 
$\mathscr{P}_1, \mathscr{P}_2, \ldots$, which are preserved both by collisions 
and by the ideal part of the dynamics.
}
\label{fig2}
\end{figure}
 
\section{Relativistic temperature}
An important question in general relativistic kinetic theory and thermodynamics concerns the translation of the classical notion of temperature into the relativistic setting. 
This problem is nontrivial, as it touches on several foundational issues that do not admit straightforward answers in relativity. 
Among these are: the formulation of relativistic thermodynamics \cite{TolmanRel}; 
the notion of thermodynamic equilibrium; 
the definition of relativistic temperature and its possible identification with a parameter of the equilibrium distribution function; 
the meaning of experimental measurement in relativity, in particular the condition of thermal equilibrium between a thermometer and a body \cite{Gavassino,Biro}; 
and the transformation law of temperature between reference frames.

The problem of temperature is already severe at the special relativistic level  \cite{Mares,Farias}, 
and originally attracted the interest of both Planck and Einstein \cite{Planck,Einstein}, who proposed the Lorentz transformation law (Planck-Einstein (PE) formula)
\begin{equation}
T = \gamma_V T', \label{vMPE}
\end{equation}
on  
how temperature values differ between two inertial frames $I$ and $I'$ in relative motion within Minkowski spacetime. Here, the temperature \( T' \) is measured in an inertial frame moving at speed \( V \) relative to another inertial frame where the body is at rest with  temperature \( T \), and appears cooler by a factor determined by the Lorentz factor \( \gamma_V = 1/{\sqrt{1 - V^2 / c^2}} \). The derivation of equation \eqref{vMPE} relies on Planck's assumption that entropy is Lorentz invariant. 

An alternative Lorentz transformation for temperature was later proposed by Ott and others, who introduced a reversible thermodynamic cycle and applied the principle of mass-energy equivalence to heat. This resulted in the Ott-Arzeli\`es (OA) formula \cite{Ott,Arzelies,Nakajima}:
\begin{equation}
T' = \gamma_V T, \label{OA}
\end{equation}
which suggests that a moving body appears hotter. The different thermodynamic assumptions underlying the discrepancy between the PE  formula and the OA formula are discussed in detail in sources such as \cite{Kampen,Nakamura}.

However, if either formula is taken to represent a general temperature transformation between inertial frames, both \eqref{vMPE} and \eqref{OA} violate the principle of equivalence of inertial frames \cite{Heras}. Specifically, they are incompatible with the special relativistic law of velocity composition \cite{Pauli}. Indeed, two successive temperature transformations, \( T \rightarrow T' \) and \( T' \rightarrow T'' \), among three inertial frames \( I \), \( I' \), and \( I'' \), should reduce to a single transformation \( T \rightarrow T'' \), where the velocity of the third frame \( I'' \) is determined by the special relativistic velocity addition law.

The above considerations imply that a universal Lorentz transformation law for temperature across inertial frames in special relativity may not be achievable, with the proper temperature (measured by a comoving observer) remaining the only consistently defined physical observable. This view has been supported by various authors, such as \cite{Landsberg,Landsberg2,Ford,Sewell}, based on the appearance of a black-body spectrum to a moving observer. It is suggested that the anisotropy introduced by the observation angle prevents the identification of a single, well-defined temperature parameter.
This issue is closely related to the measurement of temperature, which depends on the concept of thermodynamic equilibrium between the thermometer and the body. It has been argued that thermodynamic equilibrium is incompatible with relative motion: the physical process of measurement requires the thermometer and the body to be at rest relative to one another. The apparent discrepancy in the PE and OA formulas can thus be attributed to different assumptions about permissible heat transfer models between the thermometer and the body \cite{Gavassino,Biro}.

The landscape of relativistic temperature described above highlights the fundamental challenge of formulating thermodynamic laws consistently within the framework of relativity. Since thermodynamics represents the macroscopic manifestation of the microscopic dynamics of a particle ensemble, its principles should be in alignment with relativistic kinetic theory. Thus, any assumptions about the covariance of thermodynamic laws, entropy, or the application of mass-energy equivalence to heat require a kinetic foundation. 

In this section we attempt a kinetic approach to the notion of temperature for a simple gas, 
regarding it as a measure of kinetic energy, rather than interpreting it as a 
parameter characterizing an equilibrium distribution function. 
Specifically, we define a kinetic temperature \( T_{ k} \) that satisfies the following properties:

\begin{enumerate}
    \item In the classical limit \( c \rightarrow +\infty \) for a perfect gas, the kinetic temperature \( T_{ k} \) corresponds to the ensemble average of particle kinetic energy \cite{Chapman,TolmanStat}.
    
    \item In the 
    Minkowski limit $g\rightarrow\eta$, \( T_{ k} \) is  proportional to the ensemble average of the special relativistic energy \( m \gamma c^2 \).
    
    \item In the  Minkowski limit $g\rightarrow\eta$, and with suitable assumptions on the conserved quantities in particle dynamics, the thermodynamic equilibria determining \( T_{ k} \) are compatible with the Maxwell-J\"uttner distribution \cite{Cubero}.
    
    \item \( T_{ k} \) respects the principle of inertial frame equivalence and the relativistic velocity addition law in 
    the Minkowski limit $g\to\eta$. 
    
    \item For non-relativistic particle velocities in a frame \( I \), the kinetic temperature \( T_{ k} \) follows a Lorentz transformation of the form \( T'_{ k} = T'_{ k}(\gamma_V, T_{ k}) \), where \( T'_{ k} \) is the temperature measured in a moving frame \( I' \) with velocity \( V \). This transformation applies solely between the rest frame of the system (or body) and the moving frame, and is not intended to transform temperatures between two moving frames.
    
    \item \( T_{ k} \) does not, in general, satisfy a universal transformation law that depends only on the Lorentz factor \( \gamma_V \) of the moving frame. Instead, the transformation law depends on both \( \gamma_V \) and the spacetime distribution of particles.
    
    \item In the general relativistic context \cite{Frankel79}, the kinetic temperature \( T_{ k} \) for a self-gravitating system in thermodynamic equilibrium conforms to the Tolman-Ehrenfest law, which relates temperature to the norm of the Killing field generating the coordinate-time symmetry. This effect predicts that a fluid column at equilibrium will exhibit a higher temperature at the bottom and is supported by relativistic fluid mechanics \cite{Tolman1930,TolmanEhrenfest} as well as the statistical mechanical interpretation of temperature as thermal time, which associates temperature with the rate \( dt/d\tau \) of coordinate-time \( t \) relative to proper time \( \tau \) \cite{Rovelli,Rovelli2}.
    
    \item The kinetic temperature \( T_{ k} \) remains well-defined as an observable, represented as a functional of the distribution function, even outside thermodynamic equilibrium, independent of its direct measurability.
\end{enumerate}

\subsection{Thermodynamic equilibrium of a simple gas}
The objective of this section is to examine certain properties of the Landau--Einstein distribution function~\eqref{LEd} for a simple gas, by which we mean that the average interaction potential energy vanishes, $\Phi = 0$. 
To simplify the notation in the following equations 
it is convenient to shift the geodesic Hamiltonian $H$ of eq. \eqref{H1} by a constant factor,  $H\rightarrow H-mc^2/2$, i.e. $H=g^{\mu\nu}p_{\mu}p_{\nu}/2m$. 
In this case, the system is described by a thermodynamic equilibrium distribution function that is independent of the coordinate-time,
\begin{equation}
f_{\infty} =\frac{1}{Z}\, \gamma\,\delta\lr{\chi}\exp\lrc{-\beta {H}}\,\zeta\lr{p_0}.
\end{equation} 
For later convenience we rescale the function $\zeta$ by a constant factor, $\zeta\rightarrow mZ\zeta/2$, so that
\eq{
f_{\infty}=\gamma\,\delta\lr{H+\frac{1}{2}mc^2}\exp\lrc{-\beta H}\zeta\lr{p_0}
}
Here, we used the fact that, since $\Phi=0$ by hypothesis, the value of $H$ is constrained to be $-mc^2/2$ on particle orbits due to the relativistic ansatz on the constancy of the speed of light. Furthermore, $H$ and $p_0$ are not independent variables. Indeed, we may solve $H$ for $p_0^H=p_0\lr{H,\bol{x},\bol{p}}$ as follows  
\eq{
p_0^H=m\frac{-\iota+\sqrt{\iota^2-\frac{2g^{00}}{m}\lr{K-H}}}{g^{00}},\label{p0}
}
where we introduced the non-diagonal component $\iota$ of the momentum $p^0$ and the spatial kinetic energy $K$ according to 
\eq{\iota=\frac{1}{m}g^{0j}p_j,~~~~K=\frac{1}{2m}p_ig^{ij}p_j.} 
Observe that equation \eqref{p0} correctly reduces to $p_0=-m\gamma c$ in Minkowski spacetime for $H=-mc^2/2$. 
Next, 
it is also convenient to define the quantity 
\eq{
p_0^m=m\frac{-\iota+\sqrt{\iota^2-\frac{2g^{00}}{m}\lr{K+\frac{1}{2}mc^2}}}{g^{00}},\label{p0m} 
}
which corresponds to the value of $p_0^H$ when $H=-mc^2/2$. 
Performing a change of variables $\lr{\bol{x},p_0,\bol{p}}\rightarrow\lr{\bol{x},H,\bol{p}}$, and noting that
\eq{
\gamma=\frac{m\iota+g^{00}p_0}{mc},}
we find that
\eq{
f_{\infty}\,d\bol{x}dp_0d\bol{p}=f_{\infty}\frac{\p p_0^H}{\p H}\,d\bol{x}dHd\bol{p}=\frac{\lr{m\iota+g^{00}p_0^H}\,\delta\lr{H+\frac{1}{2}mc^2}\exp\lrc{-\beta H}\zeta\lr{p_0^H}}{mc\sqrt{\iota^2-\frac{2g^{00}}{m}\lr{K-H}}}\,d\bol{x}dHd\bol{p}.\label{fdH}
}
Let \eq{\mf{f}\lr{t,\bol{x},\bol{p}}=\int f\,dp_0,}
denote the distribution function on the measure $d\bol{x}d\bol{p}$. By integrating equation \eqref{fdH} with respect to $H$, we find   the steady state $\mf{f}_{\infty}\lr{\bol{x},\bol{p}}$ with expression  
\eq{
\mf{f}_{\infty}=\frac{\lr{m\iota+g^{00}p_0^m}\zeta\lr{p_0^m}}{m\sqrt{\iota^2-\frac{2g^{00}}{m}\lr{K+\frac{1}{2}mc^2}}}=\frac{c\gamma^m\zeta\lr{p_0^m}}{\sqrt{\iota^2-\frac{2g^{00}}{m}\lr{K+\frac{1}{2}mc^2}}}=
\zeta\lr{p_0^m}
,\label{feq}
}
where we defined the Lorentz factor evaluated at $p_0^m$ according to
\eq{
\gamma^m=\frac{m\iota+g^{00}p_0^m}{mc}=\frac{1}{c}\sqrt{\iota^2-\frac{2g^{00}}{m}\lr{K+\frac{1}{2}mc^2}},\label{gm}
}
and where we absorbed a constant factor $\exp\lrc{\beta mc^2/2}/c$ into the definition of $\zeta$ for  notational convenience.


\subsection{Average  spatial kinetic energy as a measure of relativistic temperature}

The aim of this section is to introduce a kinetic definition of temperature for a self-gravitating simple gas. 

First, we discuss several possible approaches that may be employed to relate temperature 
and kinetic energy in the relativistic setting.
 In classical kinetic theory, temperature is defined as the average kinetic energy of particles in a system \cite[pp. 73-79]{Huang}. To generalize this concept covariantly, we may introduce the following notion of relativistic kinetic temperature, \(T_0\), in terms of the momentum-averaged particle energy \(H\):
\begin{equation}
    \frac{4}{2} k_B T_0 = -\frac{\int H f \, dp}{\int f \, dp} = \frac{1}{2} mc^2\implies T_0=\frac{mc^2}{4k_B},
    \label{T0}
\end{equation}
where we have used the fact that the speed of light \(c\) imposes the constraint \(H = -\frac{1}{2} mc^2\) for all particles. Here, \(k_B\) is the Boltzmann constant, and the factor $4/2$ has been included to keep the analogy with the classical equipartition theorem. However, this definition, though invariant across different observers, does not reduce to the classical temperature in the limit \(c \to +\infty\). This highlights that the classical concept of temperature is associated with the spatial component of kinetic energy rather than the full spacetime dynamics.

The next candidate $T_{p_0}$ for a relativistic kinetic temperature for a system at thermodynamic equilibrium can be formulated in terms of the momentum average of the 
invariant $p_0$ associated with the coordinate-time Killing symmetry $\p_t$,
\eq{
\frac{3}{2}k_BT_{p_0}=-\frac{\int \lr{cp_0+mc^2}f\,dp}{\int f\,dp}=-c\frac{\int {p_0}f\,dp}{\int f\,dp}-4k_BT_0.
}
The temperature $T_{p_0}$ is not covariant (this is clear in Minkowski spacetime where $p_0=-m\gamma c$), and only applies to equilibrium, where $p_0$ is a constant. 
Indeed, in the absence of a coordinate-time Killing  symmetry 
$p_0$ ceases to behave as an invariant `energy', and there appears to be no special reason to associate the momentum average of the  momentum $p_0$ to a notion of temperature. 
Nevertheless, $T_{p_0}$ has the merit that it reduces to the classical temperature for small $\bol{v}/c$ and Minkowski spacetime, since in such regime \eq{
cp_0+mc^2\approx-\frac{1}{2}m\bol{v}^2,} 
where $\bol{v}=d\bol{x}/dt$. 

A third non-covariant notion $T_{ k}$ of relativistic kinetic temperature, reproducing the classical temperature in the non-relativistic limit, applicable to non-equilibrium systems, and consistent with the Tolman-Ehrenfest law, 
can be formulated by returning to the idea that temperature measures the average \ti{spatial} kinetic energy of a particle. 
In this context, the spatial kinetic energy is given by 
\begin{equation}
K= \frac{1}{2m} p_i g^{ij} p_j. 
\end{equation}
However, in the Minkowski limit, this expression does not reduce to the special relativistic energy $cp_0 = -m\gamma c^2$. This suggests that defining kinetic temperature based on the ensemble average of $K$ would lead to discrepancies with the corresponding definition of kinetic temperature in special relativity, which is based on $p_0$.

On the other hand, the fact that in the Minkowski limit the notion of temperature is tied to the momentum average of the Lorentz factor $\gamma$ (since $cp_0 = -m\gamma c^2$) suggests that the characteristic kinetic energy of the ensemble is related to the typical value of the Lorentz factor, i.e., the momentum average of the coordinate-time to proper time ratio $dt/d\tau$. This relationship is explicitly seen in equation \eqref{gm}, where the Lorentz factor $\gamma^m$ is expressed as a function of $K$, $g^{00}$, and $\iota$. 
Therefore, we are led to  the following measure of spatial kinetic energy:
\begin{equation}
k=mc^2\gamma^m=mc^2\sqrt{\frac{\iota^2}{c^2}-{g^{00}}\lr{\frac{2K}{mc^2}+1}}
.\label{kappa}
\end{equation}
Note that this form ensures that, when using the Minkowski metric, $k$ reduces to $-cp_0$, maintaining consistency with the special relativistic expression. 
Then, the relativistic kinetic temperature $T_{k}$ can be defined as the momentum average of $k$:
\begin{mydef} The relativistic temperature $T_{k}$ is defined as 
\eq{
\frac{3}{2}k_BT_{k}=\frac{\int k f\,dp}{\int f\,dp}=\frac{\int k\mf{f}\,d\bol{p}}{\int\mf{f}\,d\bol{p}}=\left\langle k\right\rangle_{\bol{p}}=mc^2\left\langle\gamma^m\right\rangle_{\bol{p}},\label{TK}
}
where   
\eq{
\langle\cdot\rangle_{\bol{p}}=\frac{1}{\rho}\int\lr{\cdot}\mf{f}\,d\bol{p},}
is the spatial momentum average and 
$\rho$ the spatial density
\eq{
\rho=\int\mf{f}\,d\bol{p}.
}
\end{mydef}
The temperature $T_{k}$ naturally reduces to the classical temperature in the non-relativistic regime, since in this limit, $k$ behaves as $mc^2\left(1 + \frac{K}{mc^2}\right)$. Moreover, $T_{k}$ remains physically meaningful far from equilibrium, as $K$ continues to represent the spatial kinetic energy of a particle. Additionally, at equilibrium, $T_{k}$ correctly reproduces the Tolman-Ehrenfest law, which relates the equilibrium temperature to the norm of the Killing field $\p_0$.
 To see this, 
observe that, from equation \eqref{gm}, we 
have
\eq{
mc^2\gamma^m=c\lr{m\iota+g^{00}p_0^m}=mc^2\sqrt{\frac{\iota^2}{c^2}-g^{00}\lr{\frac{2K}{mc^2}+1}}.
}
It follows that
\eq{
\frac{3}{2}k_BT_{k}=&c\frac{\int\lr{m\iota+g^{00}p_0^m}\mf{f}\,d\bol{p}}{\int\mf{f}\,d\bol{p}}
.\label{TK2}
}
The expression \eqref{TK2} can be cast in the simpler form 
\eq{
\frac{3}{2}k_BT_{k}=c\left\langle
{m\iota +g^{00}p_0^m}
\right\rangle_{\bol{p}}.\label{TK3}
}
Now consider the special case in which the spacetime part of the metric tensor 
vanishes, $g^{0i}=0$, so that 
$\iota=0$, and the system is at equilibrium with 
steady distribution $\mf{f}_{\infty}=\zeta\lr{p_0^m}$ given by equation \eqref{feq} for a self-gravitating simple gas.  
Then, equation \eqref{TK3} reduces to
\eq{
\frac{3}{2}k_BT_{k}=cg^{00}\left\langle p_0^m\right\rangle_{\bol{p}}.\label{Tolman0}
}
On the other hand, in this setting equation \eqref{p0m} reduces to
\eq{
c\sqrt{-g^{00}}p_0^m=-mc^2\sqrt{1+\frac{2K}{mc^2}}.
}
Hence,
\eq{
\frac{3}{2}k_BT_{k}=mc^2\sqrt{-g^{00}}\left\langle\sqrt{1+\frac{2K}{mc^2}}\right\rangle_{\bol{p}}.
}
Now consider a weakly relativistic regime such that $2K/mc^2=O\lr{\epsilon}$ and  $g^{ij}=\eta^{ij}+\epsilon h^{ij}+O\lr{\epsilon^2}$, 
and $\bol{p}^2=p_i\eta^{ij}p_j$, where $\epsilon>0$ is  small ordering parameter. Then, at first order in $\epsilon$ we obtain 
\eq{
\frac{3}{2}k_BT_{k}
=\sqrt{-g^{00}}\lr{mc^2+\left\langle{\frac{\bol{p}^2}{2m}}\right\rangle_{\bol{p}}}.\label{Tolman}
}
Hence, if the momentum average of the leading (first order)  contribution to the spatial kinetic energy
$\langle \bol{p}^2/2m\rangle_{\bol{p}}$ is a spatial constant up to $O\lr{\epsilon^2}$ corrections, the relationship \eqref{Tolman} corresponds to 
the Tolman-Ehrenfest law 
\eq{T_{k}\abs{\p_0}={\rm constant},}
relating temperature to the norm $-g_{00}=\abs{\p_0}^2=-1/g^{00}$ of the Killing field $\p_0$ in a weak field regime. 
The spatial constancy of $\langle\bol{p}^2/2m\rangle_{\bol{p}}$ to the relevant order can be verified explicitly, for example, when $\zeta\propto\exp\lrc{\beta cp_0^m}$ as in the J\"uttner distribution. 

{As an example, let us evaluate $T_{k}$ for a Synge gas \cite{Israel}.} 
Neglecting rigid rotations and restricting to Minkowski spacetime, the equilibrium distribution function of a Synge gas takes the form:
\begin{equation}
\mf{f}_{\mathrm{Sy}}(\bol{p}) = \alpha \exp\left(\beta_{\mu} p^{\mu}\right),
\end{equation}
where \( \alpha \) and \( \beta_{\mu} \) are constants satisfying \( \beta_{\mathrm{Sy}}^2 = -\beta_{\mu} \beta^{\mu} \), and \( \beta_{\mu} = \beta_{\mathrm{Sy}} u_{\mu}/c \) with \( u_{\mu} u^{\mu} = -c^2 \). Here, \( p_0(\bol{p}) \) is determined by the mass-shell condition \( H = -mc^2/2 \). 
Using the equilibrium distribution function \( \mf{f}_{\mathrm{Sy}} \), we can relate the kinetic temperature \( T_{k} \) to the Synge temperature \( c / \beta_{\mathrm{Sy}} \):
\begin{equation}
\frac{3}{2}k_BT_{k} = \frac{\int k \mf{f}_{\mathrm{Sy}} \, d\bol{p}}{\int \mf{f}_{\mathrm{Sy}} \, d\bol{p}} = mc^2 \frac{\int \sqrt{1 + \frac{\bol{p}^2}{m^2 c^2}} \exp\left[\beta_0 mc \sqrt{1 + \frac{\bol{p}^2}{m^2 c^2}} + \beta_i p^i\right] d\bol{p}}{\int \exp\left[\beta_0 mc \sqrt{1 + \frac{\bol{p}^2}{m^2 c^2}} + \beta_i p^i\right] d\bol{p}},
\end{equation}
where \( \beta_0^2 = \beta_{\mathrm{Sy}}^2 + \beta_i \beta^i \). At first order for small \( \bol{p}/c \) and \( \beta_i = 0 \),
\begin{equation}
\frac{3}{2}k_BT_{k} = mc^2 + \frac{3c}{2\beta_{\mathrm{Sy}}}.
\end{equation}
Finally, we emphasize that the key difference between \( T_{k} \) and \( c / \beta_{\mathrm{Sy}} \) lies in their applicability. While \( c / \beta_{\mathrm{Sy}} \) is a parameter characterizing the equilibrium distribution function of a Synge gas, \( T_{k} \) serves as a measure of spatial kinetic energy and applies to collisionless systems, systems influenced by spacetime curvature during collisions, as well as systems out of equilibrium. This versatility makes \( T_{k} \) well-suited for describing astrophysical systems or magnetized plasmas that exhibit collisionless relaxation \cite{SatoMorr24,Lynden,Chavanis} or experience violations of momentum conservation during binary scatterings due to curvature effects, such as in gravitational interactions within a statistical ensemble of massive astrophysical objects.




\subsection{Lorentz transformation of relativistic temperature}

The aim of this section is to determine the Lorentz transformation of the kinetic temperature $T_{k}$ defined in the previous section.

We consider two frames, $I$ and $I'$, with coordinates $\lr{x^0,x^1,x^2,x^3} = \lr{ct, x, y, z}$ and $\lr{y^0,y^1,y^2,y^3} = \lr{ct', x', y, z}$, respectively, such that their origins coincide at $t = t' = 0$. The frame $I'$ is assumed to move with a velocity $dx/dt = V$ along the $x^1$-direction relative to $I$.

To simplify the analysis, we will restrict our calculations to the case of a diagonal metric tensor, implying $\iota = 0$. Recall that in this setup, the kinetic temperature is defined as the momentum average of $c g^{00} p_0^m$. Consequently, the difference in the temperatures $T_{k}$ and $T_{k}'$ observed in $I$ and $I'$ will be determined by the difference in the measured values of $c g^{00} p_0$ and by the distinction between the momentum averaging operators $\langle \cdot \rangle_{\bol{p}}$ and $\langle \cdot \rangle_{\bol{p}}'$.

Next, observe that 
\eq{
-\lr{g^{00}}^2p_0^2=-m^2\lr{\frac{dx^{0}}{d\tau}}^2=-m^2c^2\lr{\frac{dt}{d\tau}}^2.
}
Because both observers in $I$ and $I'$ will agree on the value of the elapsed proper time $d\tau$, our task is thus to establish the relationship between ${dt}^2$ and
$\lr{dt'}^2$. Let $g_{\mu\nu}'$ denote the metric tensor in the coordinate system $\lr{y^0,y^1,y^2,y^3}$ of the $I'$ frame. 
From the definition of $d\tau$, we have
\eq{
-c^2d\tau^2=c^2g_{00}dt^2
+g_{ij}dx^idx^j
=c^2g_{00}'\lr{dt'}^2+
g_{ij}'{dy^i}{dy^j}.
} 
For $y^0=t'$, $y^1=x'$, $y^2=x^2$, $y^3=x^3$, $g_{22}'=g_{22}$, and $g_{33}'=g_{33}$, 
it follows that
\eq{
c^2g_{00}dt^2+g_{11}dx^2=c^2g_{00}'{dt'}{}^2+g'_{11}dx'{}^2.\label{dt}
}
In order to express $dt^2$ entirely in terms of momentums observed in the $I'$ frame, we now introduce the following notation for the transformation of velocities,
\sys{
&\frac{dx}{dt}=\gamma_x\lr{\frac{dx'}{dt'}+V},\\
&\frac{dy}{dt}=\frac{1}{\gamma_t}\frac{dy}{dt'},\\
&\frac{dz}{dt}=\frac{1}{\gamma_t}\frac{dz}{dt'}.
}{vt}
Recall that in Minkowski spacetime we have
\eq{
\gamma_x=\frac{1}{1+\frac{V}{c^2}\frac{dx'}{dt'}},~~~~\frac{dt}{dt'}=\gamma_t=\frac{\gamma_V}{\gamma_x}=\frac{1+\frac{V}{c^2}\frac{dx'}{dt'}}{\sqrt{1-\frac{V^2}{c^2}}},
}
with $\gamma_V=1/\sqrt{1-V^2/c^2}$ the Lorentz factor associated with the $I'$ frame velocity $V$. 
Using equation \eqref{dt}, we thus arrive at
\eq{
c{g^{00}}p_0&=mc^2\frac{dt}{d\tau}=
mc\sqrt{-g^{00}}\sqrt{-c^2g_{00}'-g_{11}'\lr{\frac{dx'}{dt'}}^2+g_{11}\gamma_x^2\gamma_t^2\lr{V+\frac{dx'}{dt'}}^2}\frac{dt'}{d\tau}\\&\neq mc^2\frac{dt'}{d\tau}=c{g^{00}{}'}p_0'.\label{change}
}
This result suggests that the temperatures $T_{k}$ and $T_{k}'$ measured in the frames $I$ and $I'$ will generally differ. 
It is now useful to quantify this difference in the special case of the Minkowski metric $g_{\mu\nu}=g_{\mu\nu}'=\eta_{\mu\nu}$. 
Setting $v_x'=dx'/dt'$, 
equation \eqref{change} becomes
\eq{
cp_0=
cp_0'\sqrt{1-\lr{\frac{v_x'}{c}}^2+\gamma_V^2\lr{\frac{V+v_x'}{c}}^2},\label{changeM}
}
where $p_0'=-m\gamma'c$ with $\gamma'=dt'/d\tau$. 
To further simplify the analysis, let us consider
the limiting case in which particles velocity
are non-relativistic in $I'$, so that $\gamma'\approx 1$ as well as $v_x'/c\ll 1$. 
Then, at leading order, equation \eqref{changeM} leads to  
\eq{
cp_0\approx c\gamma_Vp_0'\approx -mc^2\gamma_V.
}
It follows that
\eq{
\frac{3}{2}k_BT_{k}=-c\langle p_0^m\rangle_{\bol{p}}\approx -c\gamma_V\frac{\int \mf{f}p_0^m{}'\,d\bol{p}}{\int\mf{f}\,d\bol{p}}
\approx mc^2\gamma_V,
\label{EP}
}
while
\eq{
\frac{3}{2}k_BT_{k}'=-c\left\langle p_0^m{}'\right\rangle_{\bol{p}'}= -c\frac{\int\mf{f}'p_0^m{}'d\bol{p}'}{\int\mf{f}'\,d\bol{p}'}\approx mc^2,
}
where $\mf{f}\lr{t',\bol{x}',\bol{p}'}$ denotes the distribution function in the frame $I'$. 
We have thus obtained the following Lorentz transformation for kinetic temperature:
\begin{equation}
T_{k} \approx  \gamma_V T_{k}', \label{EP2}
\end{equation}
which applies to particles with non-relativistic speeds in the frame $I'$ (which are thus approximately at rest in $I'$). 
This transformation indicates that if one measures the kinetic temperature $T_{k}$ of a particle ensemble in a moving frame $I$, the observed temperature will increase as $V$ approaches $c$. In other words, a moving system appears hotter.

It is worth noting that the Lorentz transformation \eqref{EP2} is the OA formula \eqref{OA}. However, note that the consistency with the velocity addition law is now not put in to question, because the hypothesis $\gamma'\approx 1$ essentially means that the ensemble is at rest in $I'$. Hence, \eqref{EP2} only relates the temperature in a rest frame to the temperature in a moving frame, and it cannot be used to transform temperatures between moving frames (with respect to the ensemble at rest). 
We 
also emphasize that \eqref{EP2} is an approximation of the complete temperature transformation, as relativistic effects due to the finite value of $v_x'/c$ and $\gamma'-1$  have been neglected in \eqref{EP2}.

\section{Concluding remarks}
In this study, we developed a Hamiltonian 
framework for general relativistic kinetic theory on the cotangent bundle $T^{\ast}M$ of a Lorentzian manifold. 
The centerpiece of the construction is the description of inter-particle interactions through a potential energy $V$ that modifies the geodesic Hamiltonian of an isolated particle. 
This approach is motivated by the observation that, in general, 
the components of the four-momentum $p_{\mu}$—in particular the temporal component $p_0$—cannot serve as additive invariants in a genuinely relativistic collision process, as such invariance requires Killing symmetries that are not generically present. 

The evolution of the distribution function is governed by the Landau–Einstein system, 
which preserves particle number and energy, satisfies an $H$-theorem, and 
admits the Landau–Einstein distribution \eqref{LEd} as its thermodynamic equilibrium. 
A key property of this equilibrium is that it reduces to the Maxwell–Boltzmann form in the classical limit, 
without invoking $p_0$ as an additive invariant in the relaxation process. 

The kinetic theory developed here also enables a discussion of the notion of temperature in general relativity. 
In particular, we introduced a kinetic temperature $T_{k}$, defined as a measure of the average spatial kinetic energy, and investigated its properties.

\section*{Statements and Declarations}
\subsection*{Data availability}
Data sharing not applicable to this article as no datasets were generated or analysed during the current study.

\subsection*{Funding}
The research of NS was partially supported by JSPS KAKENHI Grant
No. 25K07267, 
No.  22H04936, and No. 24K00615. 

\subsection*{Competing interests} 
The authors have no competing interests to declare that are relevant to the content of this article.

\subsection*
{Disclaimer}
This is a preprint. It has not been peer-reviewed or submitted for publication. Please contact the author before using.

\printbibliography



 



\end{document}